\documentclass[aps, prd, onecolumn, superscriptaddress, groupedaddress]{revtex4} 

\usepackage{graphicx}	
\usepackage{amssymb}
\usepackage{amsmath}
\usepackage{dcolumn}
\usepackage{color}
\usepackage{subfigure, rotating, bm, array}

\usepackage[pagebackref=false, colorlinks=true]{hyperref}
\hypersetup{
linkcolor=blue,     
citecolor=blue,     
urlcolor=blue}     

\begin{document}

\title{Optical appearance of regularized compact objects without an exterior photon sphere}
\author{Ashok B. Joshi}
\email{gen.rel.joshi@gmail.com}
\affiliation{PDPIAS,
Charotar University of Science and Technology, Anand- 388421 (Guj), India.}
\affiliation{International Centre for Space and Cosmology, School of Arts and Sciences, Ahmedabad University, Ahmedabad-380009 (Guj), India.}
\author{Vishva Patel}
\email{vishvapatelnature@gmail.com}
\affiliation{PDPIAS,
Charotar University of Science and Technology, Anand- 388421 (Guj), India.}
\affiliation{International Centre for Space and Cosmology, School of Arts and Sciences, Ahmedabad University, Ahmedabad-380009 (Guj), India.}
\author{Parth C. Varasani}
\email{varasaniparth2014@gmail.com}
\affiliation{International Centre for Space and Cosmology, School of Arts and Sciences, Ahmedabad University, Ahmedabad-380009 (Guj), India.}

\date{\today}

\begin{abstract}
Recent observations by the Event Horizon Telescope (EHT) indicate that the shadow of the compact object at our Galaxy's center (Sgr A*) closely resembles that of a Schwarzschild black hole. However, identifying the presence and exact location of unstable circular null geodesics outside the compact object (hereafter referred to as an exterior exterior photon sphere) observationally remains challenging. Motivated by this, we investigate shadow formation in spacetimes that lack an exterior photon sphere by applying the Simpson--Visser (SV) regularization technique (originally designed to smooth black hole singularities) to null singularity and charged null singularity metrics. We show that the regularized null and charged null singularity spacetimes exhibit shadow-like features despite the absence of an exterior exterior photon sphere. We analyze how the SV regularization parameter influences their geometry and shadow size, and show that the regularized null and charged null singularity spacetimes can correspond either to two-way traversable wormholes or retain singularities. For the wormhole branch, the spacetime admits an unstable null orbit at the throat while remaining free of an exterior exterior photon sphere. Apart from the shadow-like dark region without an exterior photon sphere in the charged SV spacetime, our investigation shows that, for a specific range of parameters, the shadow boundary (dark region) is controlled by the regular core rather than by an exterior photon sphere. Our results reveal that shadows arising from these regularized null singularity spacetimes closely mimic those of Schwarzschild and charged black-bounce spacetimes, even though no exterior exterior photon sphere exists. We also perform a phenomenological comparison of the predicted shadow sizes with the EHT observations of Sgr A* and M87, identifying parameter ranges compatible with the observed angular diameters.

\bigskip

$\boldsymbol{key words}$ : Shadow, Regular compact objects, Light trajectories.
\end{abstract}
\maketitle
\section{Introduction}
Recent advances in observational astrophysics, most notably the Event Horizon Telescope (EHT) collaboration's imaging of the M87 galactic center \cite{EventHorizonTelescope:2019dse, EventHorizonTelescope:2019uob, EventHorizonTelescope:2019jan, EventHorizonTelescope:2019ths, EventHorizonTelescope:2019pgp, EventHorizonTelescope:2019ggy, EventHorizonTelescope:2020qrl}, have opened a new window to analyze the nature of compact objects. The direct images suggest that the shadow size of the central supermassive object, Sgr A*, closely resembles that predicted for a Schwarzschild black hole \cite{EventHorizonTelescope:2022xnr, EventHorizonTelescope:2022vjs, EventHorizonTelescope:2022wok, EventHorizonTelescope:2022exc,
EventHorizonTelescope:2022urf, EventHorizonTelescope:2022xqj, 
EventHorizonTelescope:2022gsd, EventHorizonTelescope:2022ago, EventHorizonTelescope:2022tzy}. The photon ring is the bright halo of light seen in black hole images, such as those from the EHT results. It is formed by photons that lensed very close to the maxima of the potential, sometimes orbiting the compact object more than once before reaching our telescopes. However, observationally locating the precise location and properties of the exterior photon sphere remains a challenge \cite{Eichhorn:2022oma}. The EHT group has yet to directly resolve the exterior photon sphere, which is still a theoretical feature. The photon ring, a lensed emission structure whose apparent radius is determined by the critical impact parameter ($b_c = 3\sqrt{3}M \approx 2.6R_s$ for a Schwarzschild black hole), has been identified through EHT observations. Although this feature does not correspond to the exterior photon sphere itself, it is in agreement with theoretical predictions and offers a useful observational probe of the strong-gravity regime near black holes \cite{Abdikamalov:2019ztb,Broderick:2024vjp,atamurotov_2015,Vagnozzi:2019apd,Gyulchev:2019tvk,Narayan,Gralla:2019xty,abdujabbarov_2015b,Li:2021,KumarWalia:2024omf,Salehi:2024cim,Ovgun:2025stp,Pedrotti:2025idg,Nalui:2025wiw,HassanPuttasiddappa:2025tji,Pantig:2024lpg,Bambi:2019tjh,Jusufi:2021lei,Solanki:2022glc,Mishra:2019trb,Ghosh:2023kge,Banerjee:2019nnj,Banerjee:2022jog,Cardoso:2014sna,Chen:2022kzv,Ozel:2021ayr,Qiao:2025ojr,Faggert:2025eja,Suzuki:2025ptr,Uniyal:2022vdu,Uniyal:2025uvc,Saurabh:2020zqg,EventHorizonTelescope:2025uqi}.\\

From the observations, researchers are still trying to figure out the position of the first subring $(n=1)$, and the thickness of the first subring will provide information on the presence of the photon ring \cite{Tiede:2022grp,Broderick:2022tfu,Lockhart:2022rui}. It is quite difficult to find direct results that determine the position of the exterior photon sphere. However, the difference between the two conjugate subrings provides the convergence rate. Logarithmic and non-logarithmic falls provide information about the exterior photon sphere \cite{lensing1}. Hence lensing in different spacetime structures provides the causal structure around the supermassive compact objects \cite{Virbhadra:2022iiy,Virbhadra:2008ws,Virbhadra:2002ju,Virbhadra:1999nm,Bozza:2010xqn,Chen:2023uuy}.\\

Examining the observational signatures of singularity nature and regularity in the context of the EHT collaboration's recent shadow investigation of the M87 galactic centre is crucial \cite{Khodadi:2020gns, Saurabh:2022jjv, Chen:2022nbb, Vagnozzi:2022moj, Khodadi:2022pqh, Khodadi:2021gbc, KumarWalia:2022ddq}. Regarding our own galaxy, the GRAVITY and SINFONI collaboration's observations of the paths of ``S" stars around our galactic center are providing incredibly valuable information. The behavior of timelike and null geodesics around singularities and other compact objects has been the subject of numerous recent investigations \cite{Eisenhauer:2005cv,center1,Wu:2023wld,Deng:2020yfm,lensing2,lensing3, Patel:2022vlu, lensing6, Gao:2020wjz, Sajadi:2023ybm}. The first work on a shadow without a exterior photon sphere was discussed in \cite{Joshi:2020tlq}. The generic end state of gravitational collapse gives the spacetime a null singularity \cite{Joshi:2023ugm}. The shadow cast by the Schwarzschild black hole is approximately five times larger than the shadow cast by the null singularity spacetime. Later work on shadows without a exterior photon sphere is explored in various articles \cite{Dey:2020bgo,Kaur,Khodadi:2024ubi}. Investigating the shadows cast by a charged null singularity shows that shadows will not form in that spacetime \cite{Viththani:2024fod}. Several studies investigated, accretion disk properties \cite{Uniyal:2025hik,Tahelyani:2022uxw,Olivares-Sanchez:2024dfh,Pugliese:2024bhh,Patra:2023epx,Prada-Mendez:2023psi,Pugliese:2023kwp,Boshkayev:2022vlv}, exterior photon spheres, which are also present in naked singularities, as well as the shadow properties in modified gravity \cite{Gan:2021xdl, Hu:2020usx, Huang:2024bbs,Feng:2022evy,DiFilippo:2024poc,Bambi:2008jg,Bambi:2017iyh}, gravastars \cite{grava2, grava3}, and wormholes \cite{bambi, ohgami:2015, Huang:2023yqd, Shaikh:2022ivr, Zhu:2021tgb,Mustapha1}. \\

This gives rise to a fundamental question: what would the observed geometry around a compact object be if a exterior photon sphere were absent? Although traditional black hole spacetimes, such as Schwarzschild, Reissner-Nordström, and Kerr metrics, possess exterior photon spheres that define their characteristic shadow. In \cite{Cunha:2020azh,Ghosh:2021txu}, given a proof that black holes at least have a one light ring (LR). Recent studies indicate that shadows may form even in spacetimes lacking such exterior photon spheres \cite{Joshi:2020tlq}. For instance, thin matter shells or certain exotic compact objects can produce shadows due to their unique light-trapping properties. These findings suggest that the exterior photon sphere is not an exclusive marker of shadow but may also appear around compact objects without the exterior photon sphere, like horizonless objects, including naked singularities. Motivated by these insights, in this work we explore the nature of shadows cast by a class of regularized spacetimes that inherently lack exterior photon spheres. We extend the Simpson-Visser (SV) regularization technique (originally designed to smooth black hole singularities) to null singularity and charged null singularity metrics.\\

It should be noted that the shadow is caused by the presence of the exterior photon sphere in the Schwarzschild, JNW, and JMN1 spacetimes \cite{Solanki:2021mkt,Shaikh:2019hbm}.  However, for a certain parameter, JNW and JMN1 can cast a shadow without a exterior photon sphere, in JNW $n = 1/2$ and in JMN1 $M_{0}=2/3$. These findings highlight the fact that other compact objects in the presence of a exterior photon sphere or thin shell of matter can also cast a shadow, demonstrating that it is not just a feature of black holes.\\

In the Simpson--Visser construction, the central curvature singularity is replaced by a finite-area minimal surface (or bounce). Depending on the value of the regularization parameter, the resulting spacetime may describe a regular black hole, a null throat, or a two-way traversable wormhole. Consequently, although the construction originates from a given seed metric, the resulting causal structure and optical properties differ significantly from those of the original spacetime.\\

In this paper, we would like to investigate the shadow in a regular black hole and other regular compact objects. The regularized black hole was initially proposed by Bardeen \cite{Bardeen} and further work has been carried out by Bardeen \cite{Bardeen:2014uaa}, Roman–Bergmann \cite{Roman:1983zza}, Frolov \cite{Frolov:2016pav}, and Hayward \cite{Hayward:2005gi}. Shadow properties in regular static and rotating regular black hole is discussed in \cite{stuchlik_2019,Bambhaniya:2021ugr,Olmo:2023lil,Boshkayev:2023fft,Kumar:2020ltt,Bambi:2013ufa,Mustapha2}. Shadows in regularizing the JNW and JMN naked singularities discussed in \cite{Pal:2022cxb,Pal:2023wqg,Pal:2024kng}. Unlike the regularized JNW and JMN spacetimes studied, we investigate a different class of asymptotically flat modified null-singularity spacetimes and demonstrate that, the shadow boundary is determined by the regular core instead of an exterior photon sphere. We further extend the analysis to the charged counterpart and compare its optical appearance with the corresponding regularized black-bounce geometries. In traditional regular black hole, Simpson and Visser proposed a different method in which the behavior of the compact object depends on the parameter $L$, i.e., depending on $L$, it is either a regular black hole or a traversable wormhole. Our key focus is on null singularity metric that, intriguingly, admit shadow formation despite the absence of exterior photon sphere. By introducing a regularization parameter \( L \), the SV approach replaces the singular center with a regular core or a traversable wormhole throat, modifying lightlike geodesics and potentially altering observational signatures. We investigate how this parameter influences the shadow size and how the resulting shadows compare with those from Schwarzschild and charged black hole spacetimes, as well as from null singularity spacetime. \\

Although several recent studies have investigated the optical appearance of Simpson--Visser regularized Schwarzschild, JNW and JMN spacetimes, the corresponding analysis for the regularized null-singularity and charged null-singularity metrics has not been explored. These geometries possess different causal structures and null geodesic properties, making it important to examine whether the shadow-like optical features found in other regularized compact objects remain valid in this distinct class of spacetimes. The primary objective of the present work is therefore to investigate the optical signatures of these geometries and to compare them with previously studied regularized compact-object solutions.\\

The Simpson–Visser construction replaces the central singularity by a finite-area minimal surface characterized by the regularization parameter L. Depending on the value of L and the metric parameters, the resulting regularized geometry may correspond to a regular black hole, a one-way null throat, or a two-way traversable wormhole. Throughout this work, we investigate the optical properties of these regularized geometries rather than those of the original singular spacetimes.\\

In this work, we apply the regularization technique to null-singularity and charged null-singularity spacetimes that Simpson and Visser used. Black-bounce-Schwarzschild spacetime and Black-bounce-Reissner- Nordström geometry are asymptotically resembling to both. We demonstrate that, despite lacking a exterior photon sphere or any thin shell of matter, this regularised spacetime cast a shadow. The regular center acts like a ``quantum object" \cite{Frolov:2021vbg,Frolov:1988vj,Modesto:2004xx,Ashtekar:2005cj,Dymnikova:1992ux,Bonanno:2000ep,Joshi:2025ozt}, with a zero-throat wormhole solution when it cannot be concealed by an event horizon in its own right, even in its ability to cast a shadow, much like a black hole or other compact objects. We consider spacetime with two free parameters: one is the mass (total mass within asymptotically flat spacetime), and the second is the regularisation parameter $L$ (SV parameter). Throughout this work, we use the term exterior exterior photon sphere to denote an unstable circular null geodesic located away from the wormhole throat. Accordingly, the unstable null orbit at the throat is referred to as a throat light ring and is distinguished from an exterior exterior photon sphere. In addition to that we employ a simplified emissivity model to investigate the qualitative influence of the spacetime geometry on optical observables. We do not attempt a full astrophysical modeling of the accretion flow or a direct comparison with EHT imaging pipelines. We distinguish between the exterior photon sphere, the bright photon ring, and the shadow boundary. The exterior photon sphere is a geometric property of the spacetime, corresponding to unstable circular null geodesics outside the regularized core. The bright photon ring is an image feature produced by highly lensed photons propagating near such unstable orbits. The shadow boundary denotes the apparent edge of the central dark region in the observed image and generally depends on the spacetime geometry, the emission model, and the adopted boundary conditions. While these concepts nearly coincide for Schwarzschild black holes, they need not coincide for horizonless regularized compact objects. A detailed reconstruction of the matter sector supporting these regularized geometries, including a systematic analysis of the associated energy conditions, as well as investigations of their dynamical formation and perturbative stability, are beyond the scope of the present work. These important aspects are left for future investigations.\\

The structure of the paper is as follows. In section (\ref{sec1}), we discuss the black bounce Schwarzschild spacetime and black bounce Reissner-Nordström geometry in subsection (\ref{sec1a}), and in subsection (\ref{sec1b}), we discuss the regularized null singularity and charged null singularity spacetimes. In section (\ref{sec3}), we discuss the nature of potential and nature of light trajectories when a exterior photon sphere is absent around the black bounce Schwarzschild and black bounce Reissner-Nordström geometries and the modified null singularity and modified charged null singularity solutions presented here. We also discuss the comparative study of the shadow in the regular black hole and regular null singularity spacetime discussed here. In section (\ref{sec4}), we discuss the radially infalling thin accretion model for intensity distribution in regular spacetime. Using observational data of Sgr A* and M87, we constrain the parameters. Finally, in section (\ref{result}), we discuss our results. Throughout the paper, we take $G=c=M=1$.

\section{SV-like deformation in black hole and naked singularity spacetimes}\label{sec1}
In this section, we discuss spherically symmetric, asymptotically flat, charged, and uncharged modified black hole and null singularity spacetimes. Using SV method in the spacetime, the metric can be given as,
\begin{equation}
    ds^2=-f(r)dt^2+\frac{dr^2}{f(r)}+(r^2+L^2)(d\theta^2+\sin^2\theta d\phi^2)
\end{equation}
where $f(r)$ is the metric component. We first discuss the basic spacetime properties of this spacetime. Here, consider the metric to be asymptotically flat, and when charge and mass vanish, it becomes the standard Morris-Thorne wormhole \cite{Morris:1988cz}. In a regular black hole, generally, null energy conditions are violated. Hence, here we have not focused on the energy condition, but we show that if a shadow without a exterior photon sphere is observed, geometry could be mentioned in this paper.

\subsection{Black-bounce and charged black-bounce spacetime}\label{sec1a}
The metric component of a black-bounce spacetime is given as follows \cite{Simpson:2018tsi};    
\begin{equation}
    f(r)=1- \frac{2M}{\sqrt{r^2+L^2}}.
\end{equation}
There are three cases possible: i) For $L>2M$, coordinate speed of light is non-zero, i.e. $|\frac{dr}{dt}|\not=0$, so it is a two-way traversable wormhole for any value of $r$. ii) For $L=2M$, and for the limit $r \to 0$ from any side, we have $|\frac{dr}{dt}|\to 0$, which shows a horizon at $r=0$, but still there is no singularity, and it resembles the one-way wormhole with an extremal null throat. iii) For $L<2M$, the coordinate speed of light $|\frac{dr}{dt}|= 0$ at horizon. There exists an event horizon at $r_{e}= \sqrt{4M^2-L^2}$. Due to this parameter $L$, the singularity has been replaced by a regular spacelike hypersurface which represents ``bounce". All other curvature tensors remain finite in this SV-spacetime, although this SV-spacetime violates the energy conditions. Also, the SV solution can be obtained by introducing a source such as a coupled field of non-linear electrodynamics with a phantom scalar field. \\

A regularised charged Black-Hole or Black-Bounce-Reissner-Nordström spacetime or charged Black-bounce spacetime is given as \cite{Franzin:2021vnj},
\begin{equation}
    f(r)=1-\frac{2M}{\sqrt{r^2+L^2}}+\frac{Q^2}{r^2+L^2}.
\end{equation}
After regularisation, the domain of $r$ has been expanded from $r \in [0, +\infty)$ to $r \in (-\infty,+\infty)$ and it preserves the asymptotic flatness.
Here $L \to 0$ represents the standard Reissner-Nordström metric, and $M\to0$ and $Q\to 0$ represent the standard Morris-Thorne wormhole. All curvature tensors and curvature scalars remain finite in the limit $r \to 0$, and violate null energy conditions. There exist horizons at $r_{\pm}= \sqrt{2M^2-L^2-q^2 \pm 2M\sqrt{M^2-q^2}}$ and, event and Cauchy horizons are absent when $q\geq M$.

\subsection{Modified charged and non-charged null singularity spacetime}\label{sec1b}
\subsubsection{Modified null singularity spacetime}
As an example of using the SV method to regularise naked singularities, we study the SV-modified version of the null singularity metric. Null singularity metric is an end state result of a realistic gravitational collapse model \cite{Joshi:2023ugm}. The  null singularity metric \cite{Joshi:2020tlq} is a static, spherically symmetric solution of the Einstein field equations for a specific matter field distribution, and is given by the line element,
\begin{equation}\label{JNW}
	ds^2=-\Big(1+\frac{M}{r}\Big)^{-2} \text{d}t^2 + \Big(1+\frac{M}{r}\Big)^{2} \text{d}r^2 + r^2 \text{d}\Omega^2~.
\end{equation}
The parameter $M$  is the ADM mass of the spacetime. The metric resembles the Schwarzschild spacetime in the first order of expansion, and in weak gravity, it behaves as Newtonian gravity. Now we apply the SV method via the replacement $r\to\sqrt{r^2+L^2}$, without modifying the form $\text{d}r$ in the metric, and the resulting metric components become
\begin{equation}
    f(r) = \Big(1+\frac{M}{\sqrt{r^2+L^2}}\Big)^{-2}. \label{nullregular}
\end{equation}
Here, $L$, the SV parameter, is a real positive quantity having dimensions of length. Also, it can readily be seen that the original symmetries and the asymptotic properties of the null singularity metric are preserved in the new one. We will get back the null singularity metric in the limit $L \rightarrow 0$. The coordinate speed of light $|\frac{dr}{dt}| \ne 0 $ for this case. The geometry describes a regular two way traversable wormhole with a minimal radius throat $L$ at $r=0$. The causal structure is governed by the locations of horizons, determined by the roots of $f(r)=0$. This condition does not yield physically valid solutions for the metric (\ref{nullregular}). Hence, the event horizon is absent in the given spacetime.\\

The definitive test of a geometry's regularity is the behaviour of its curvature invariants. For the metric (\ref{nullregular}), we have verified that the full expressions for the Ricci scalar R(r) and the Kretschmann scalar K(r) remain finite for all finite values of r provided $L > 0$. In particular, evaluating these invariants near the center, as r$\to$ 0, we obtain,

\begin{equation}
    R \to -\frac{2(-L M^4+L^4(L+M)-L^2(3LM^2+4M^3))}{L^3 (L+M)^4},
\end{equation}
which is finite. More comprehensively, the Kretschmann scalar $K = R_{\mu\nu\rho\sigma}R^{\mu\nu\rho\sigma}$ is also finite as $r \to 0$,
\begin{equation}
     K \to 4\bigg( \frac{2 L^4(L+M)^4+L^2M^2(L^2+ML)^2}{L^4(L+M)^8} + \frac{(L+M)^4(L^2+M(2L+M)^2)}{L^4(L+M)^8}\bigg),
\end{equation}
   
Thus, the curvature invariants remain bounded throughout the spacetime for L $>$ 0, confirming that the Simpson-Visser regularization removes the curvature singularity present in the original metric. The finite limits at r$\to$0 shown above explicitly illustrate the regular behaviour at the spacetime center. In the asymptotic limit r$\to \infty$, the spacetime is asymptotically flat.

Here, we analyse energy conditions in the Einstein field.
The physical nature of the supporting matter is best analyzed by the energy conditions. The Null Energy Condition (NEC), which states $T_{\mu\nu}k^\mu k^\nu \ge 0$ for any null vector $k^\mu$, is the most fundamental of these. For a spherically symmetric fluid, this requires $\rho+p_r \ge 0$ and $\rho+p_t \ge 0$. Using the Einstein tensor from the \cite{Simpson:2018tsi}, we find:

\begin{equation}
    \rho + p_r = \frac{1}{8\pi} (-T^t_{\ t} + T^r_{\ r}) 
 =- \frac{L^2}{4\pi(L^2+r^2)(L^2+M^2+2M\sqrt{L^2+r^2}) }.
\label{eq:nec_radial}
\end{equation}
This term is negative, indicating a clear violation of the NEC. This violation is not an incidental feature but a requirement that singularity resolution within classical general relativity is seem difficult without considering the presence of such ``exotic'' matter.

\subsubsection{Modified charged null singularity spacetime} 
The charged null singularity metric \cite{Viththani:2024fod} is a static, spherically symmetric solution of the Einstein-Maxwell field equations. The line element of spacetime is given as follows:
\begin{equation}
    f(r) = \Big(1+\frac{M}{r}\Big)^{-2} +\frac{q^2}{r^{2}}. 
\end{equation}
The parameters $M$ and $q$ are the ADM mass and the charge of the spacetime. In the weak field limit, this metric resembles the RN spacetime. Now we apply the SV method via the replacement $r\to\sqrt{r^2+L^2}$, without modifying the form $\text{d}r$ in the metric, and the resulting metric becomes, 
\begin{equation}
    f(r) = \Big(1+\frac{M}{\sqrt{r^2+L^2}}\Big)^{-2} +\frac{q^2}{(r^2 +L^2)}. \label{chargeregular} 
\end{equation}
It can easily be verified that the original symmetries and asymptotic behaviour of the null singularity metric are maintained in the new metric. In the limit $L \rightarrow 0$, the charged null singularity metric is recovered. The coordinate speed of light $|\frac{dr}{dt}| \ne 0 $ for this case. The geometry describes a regular, two-way traversable wormhole with a throat of minimal radius $L$ at $r=0$.\\

Now, analysing the behaviour of curvature scalars for the metric (\ref{chargeregular}), by providing $L>0$. The Ricci scalar as $r \to 0$ is calculated to be:
{\small
\begin{equation}
\begin{split}
R\to \frac{-2}{L^7 (L + M)^4} \times \Biggl(
    & L^8 (L + M) 
    + L^3 M^4 q^2 
     + L^6 \bigl( -3 L M^2 - 4 M^3 + L q^2 + 4 M q^2 \bigr) \\
    & + L^4 \bigl( -L M^4 + 6 L M^2 q^2 + 4 M^3 q^2 \bigr)
\Biggr)
\end{split}
\end{equation}
}
Which is finite, and also the Kretschmann scalar  $K = R_{\mu\nu\rho\sigma}R^{\mu\nu\rho\sigma}$ remains finite as $r \to 0$, which resembles the absence of curvature singularity, and as $r \to \infty$ this recovers asymptotically flat spacetime. 
The causal structure is governed by the locations of horizons, determined by the roots of $f(r)=0$. This condition yields, with no real solution existing for $r$, and hence no horizon.\\

For a specific kind of source, the metric is a solution to the Einstein field equations, $G_{\mu\nu}=8\pi T_{\mu\nu}$. To analyse the nature of this kind of matter, one can use the energy conditions, specifically the Null Energy Condition (NEC), i.e. $T_{\mu\nu}k^{\mu}k^{\nu}\ge 0$ for any null vector $k^{\mu}$. For a spherically symmetric fluid it requires $\rho + p_{r} \ge 0$ and $\rho + p_{t} \ge 0$. From this, we found that this condition does not hold, and it signifies some exotic matter present in this kind of spacetime.

\begin{figure*}
\centering
\subfigure[The effective potential in SV spacetime.]
{\includegraphics[width=62mm]{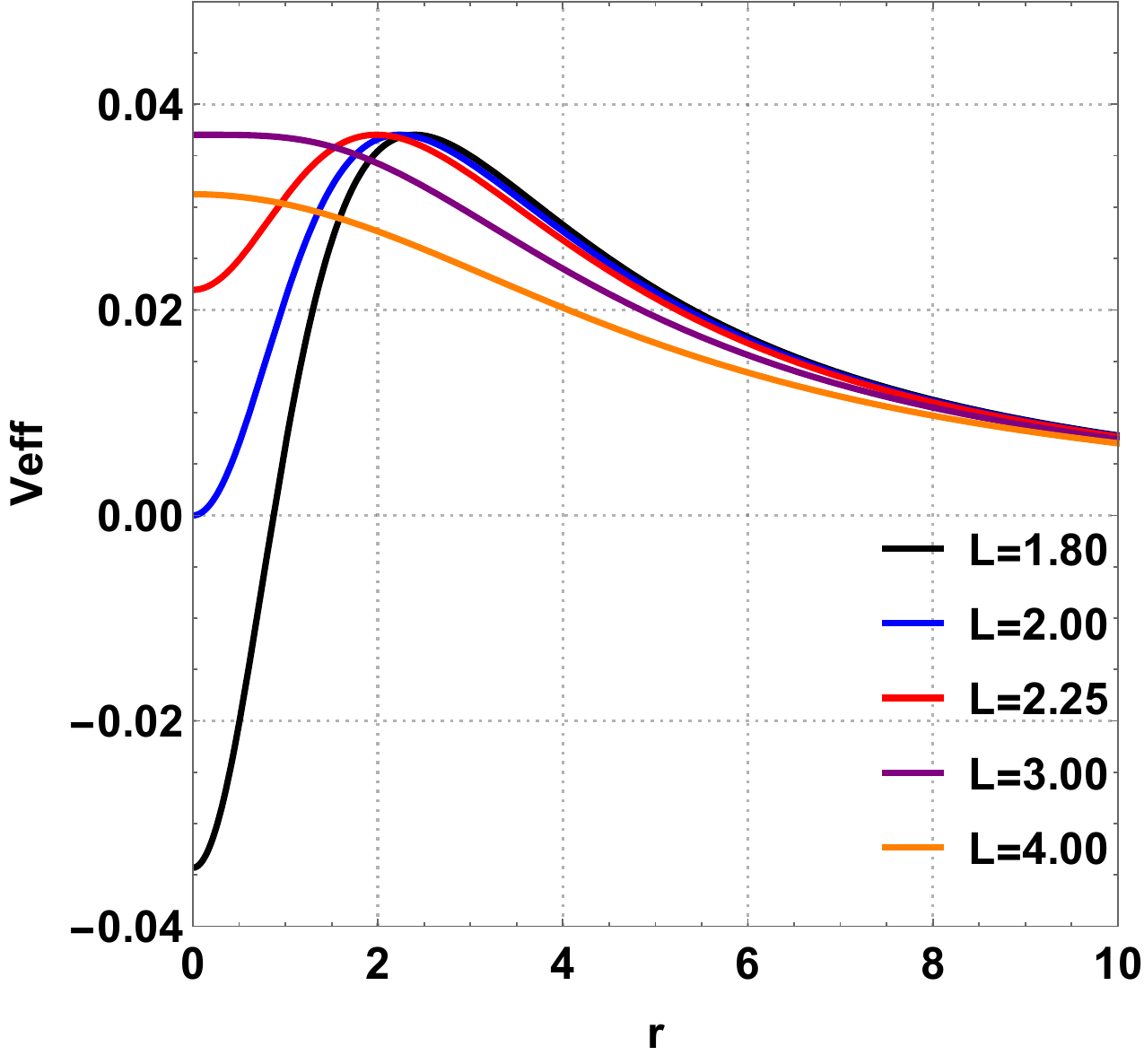}\label{fig1a}}
\hspace{1.2cm}
\subfigure[Lensing effect in SV spacetime.]
{\includegraphics[width=58mm]{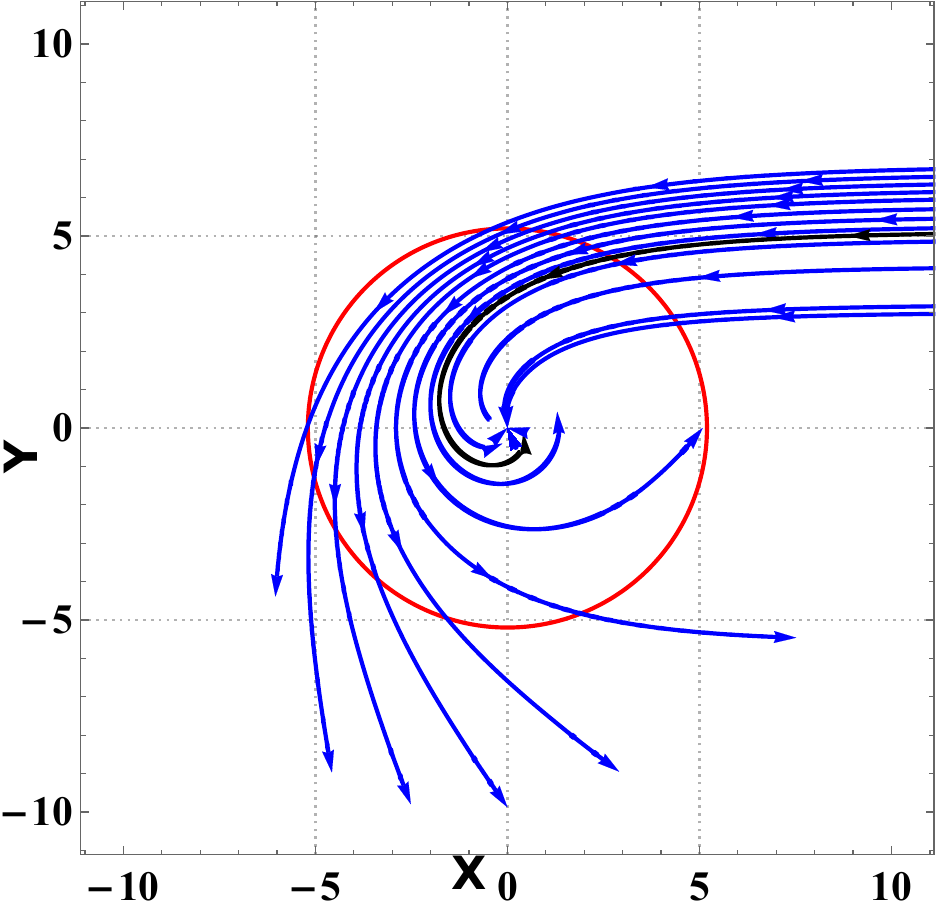}\label{fig1b}}\\
\hspace{0.2cm}
\subfigure[Intensity distribution in SV spacetime.]
{\includegraphics[width=82mm]{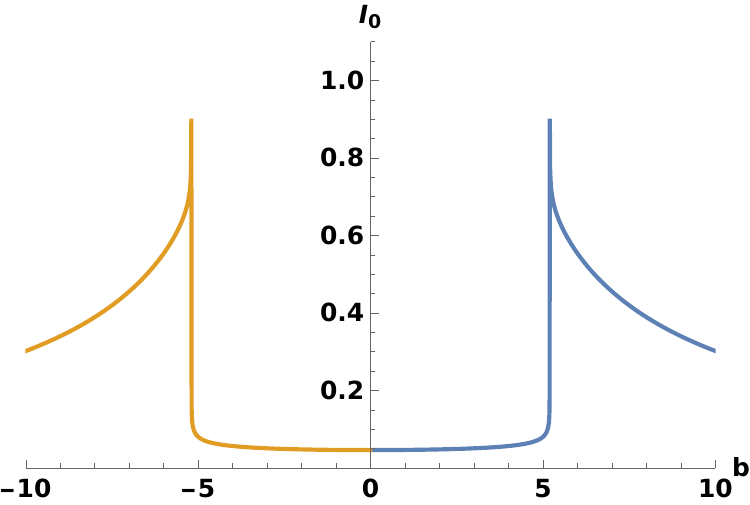}\label{fig1c}}
\hspace{0.2cm}
\subfigure[Shadow in SV spacetime.]
{\includegraphics[width=60mm]{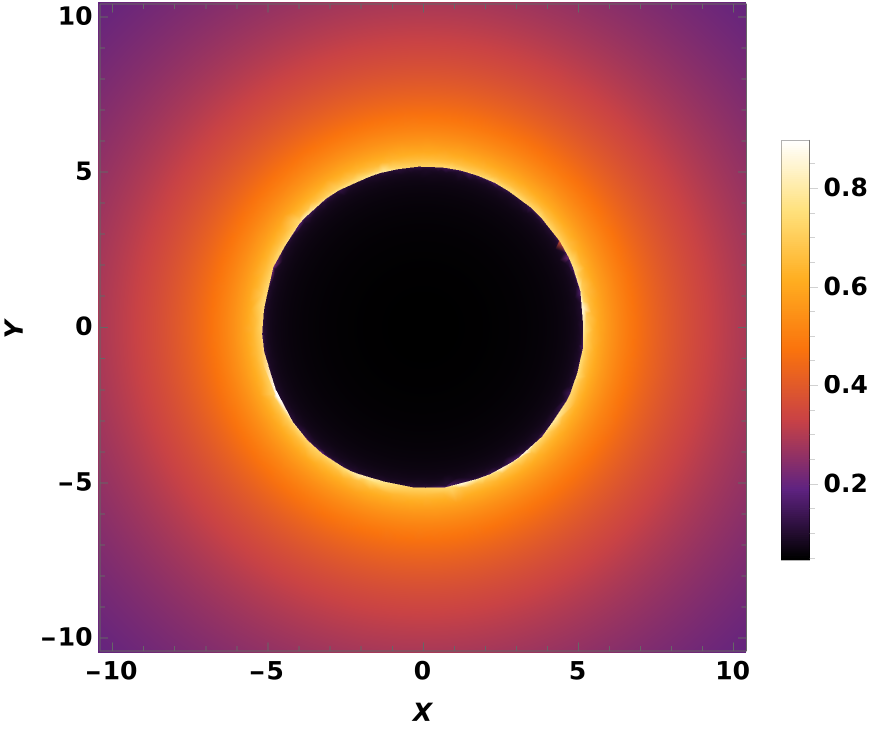}\label{fig1d}}
 \caption{In Figs.~(\ref{fig1a}), (\ref{fig1b}), (\ref{fig1c}), and (\ref{fig1d}) we show the nature of effective potentials of the null geodesics, light trajectory, intensity distribution, and shadow image in black-bounce (SV) spacetime. The effective potential of light-like geodesics for different impact parameters in the black-bounce spacetime is shown in Fig.~(\ref{fig1a}). In Fig.~(\ref{fig1b}), the light trajectories in these spacetimes are shown, and the blue lines are the null geodesics. In Figs.~(\ref{fig1c}) and (\ref{fig1d}), the intensity map in observer sky and the shadow of the central object are shown for the black-bounce geometry. The shadow shown in the right bottom corner is the shadow cast by the black-bounce spacetime.}\label{fig1}
\end{figure*}

\begin{figure*}
\centering
\subfigure[The effective potential in charged SV spacetime.]
{\includegraphics[width=62mm]{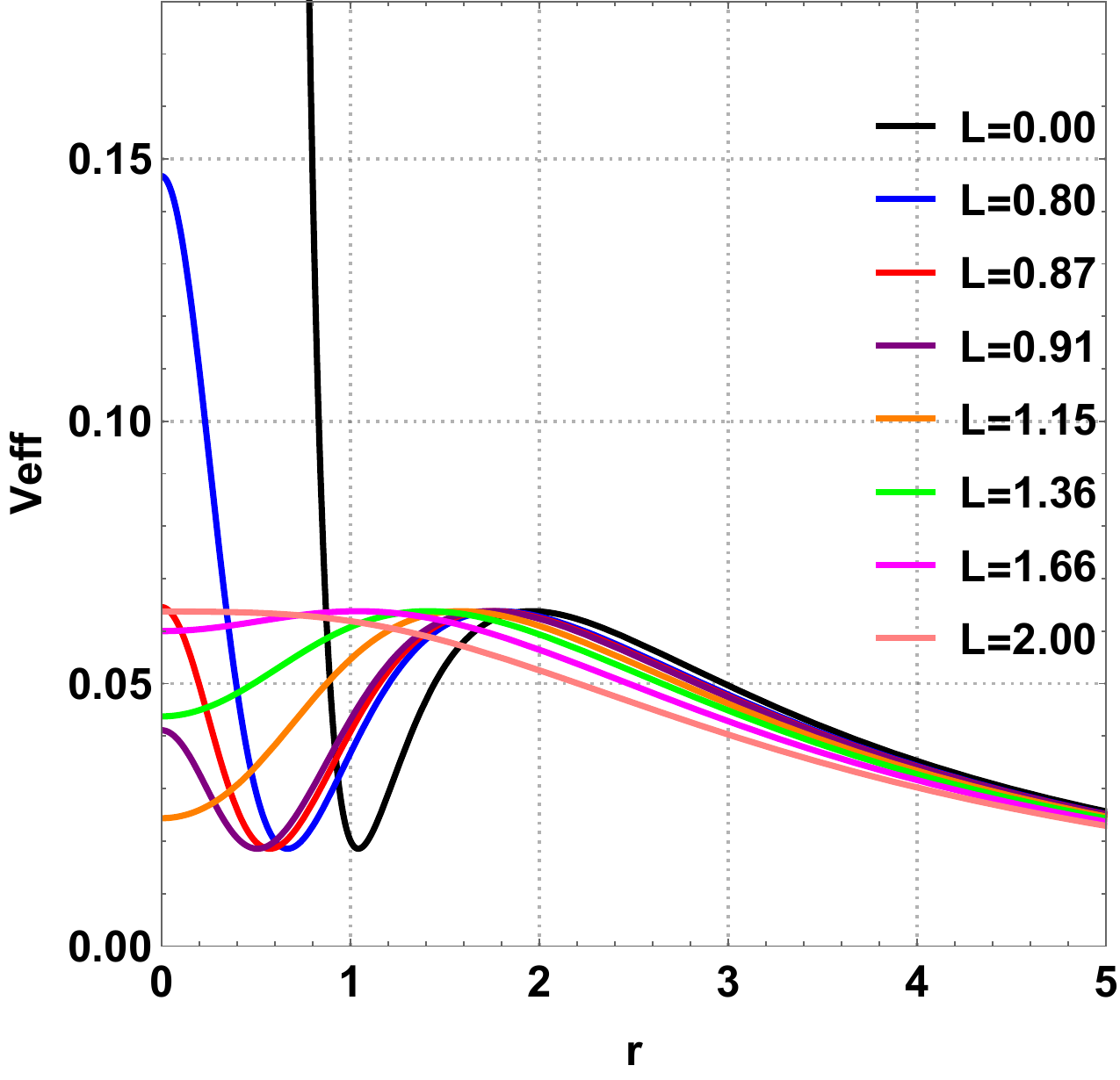}\label{fig2a}}
\hspace{1.2cm}
\subfigure[Lensing effect in charged SV spacetime.]
{\includegraphics[width=58mm]{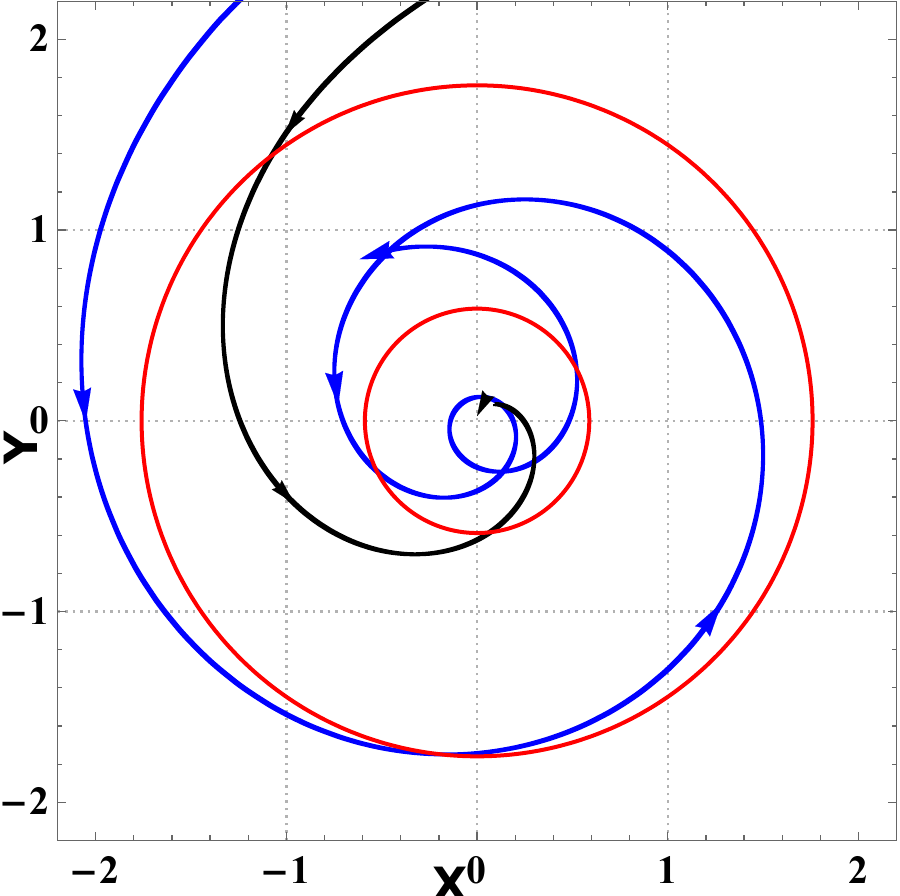}\label{fig2b}}\\
\hspace{0.2cm}
\subfigure[Intensity distribution in charged SV spacetime.]
{\includegraphics[width=82mm]{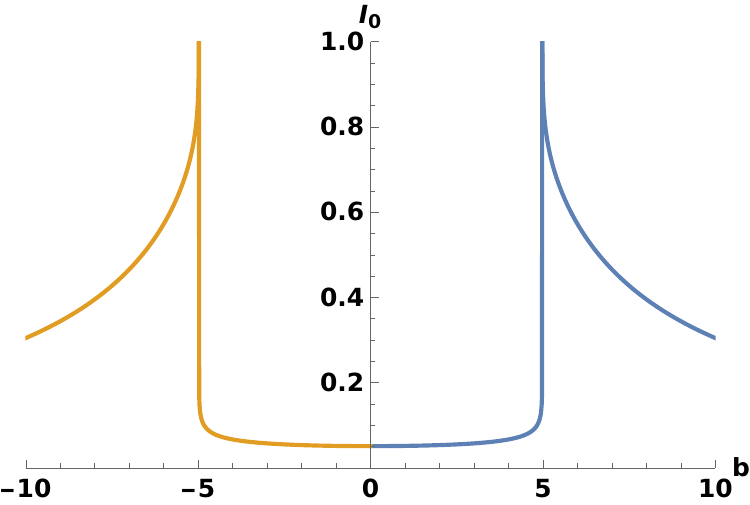}\label{fig2c}}
\hspace{0.2cm}
\subfigure[Shadow in charged SV spacetime.]
{\includegraphics[width=68mm]{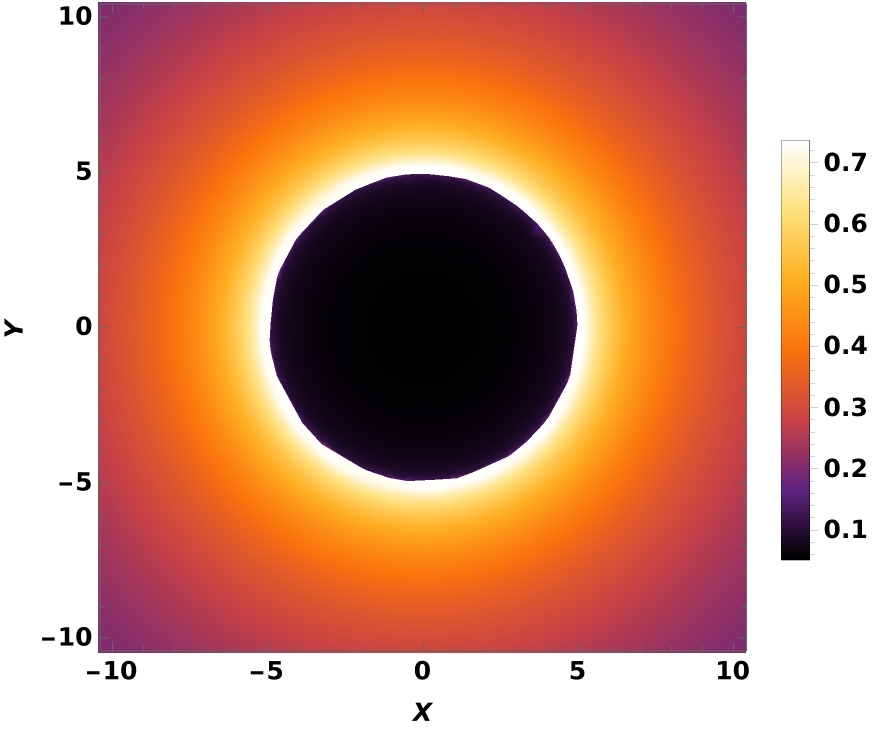}\label{fig2d}}
 \caption{In Figs.~(\ref{fig2a}), (\ref{fig2b}), (\ref{fig2c}), and (\ref{fig2d}) we show the nature of effective potentials of the null geodesics, light trajectory, intensity distribution, and shadow image in charged black-bounce (SV) spacetime. Effective potential of light-like geodesics for different impact parameter in charged black-bounce spacetime is shown in Fig.~(\ref{fig2a}). In Fig.~(\ref{fig2b}) the light trajectories in these spacetimes are shown, the blue lines are the null geodesics. In Figs.~(\ref{fig2c}) and (\ref{fig2d}), the intensity map in observer sky and the shadow of the central object are shown for charged black-bounce spacetime. The shadow shown in the right bottom corner is the shadow cast by the charged black-bounce spacetime.}\label{fig2}
\end{figure*}

\begin{figure*}
\centering
\subfigure[The effective potential in modified null singularity spacetime.]
{\includegraphics[width=68mm]{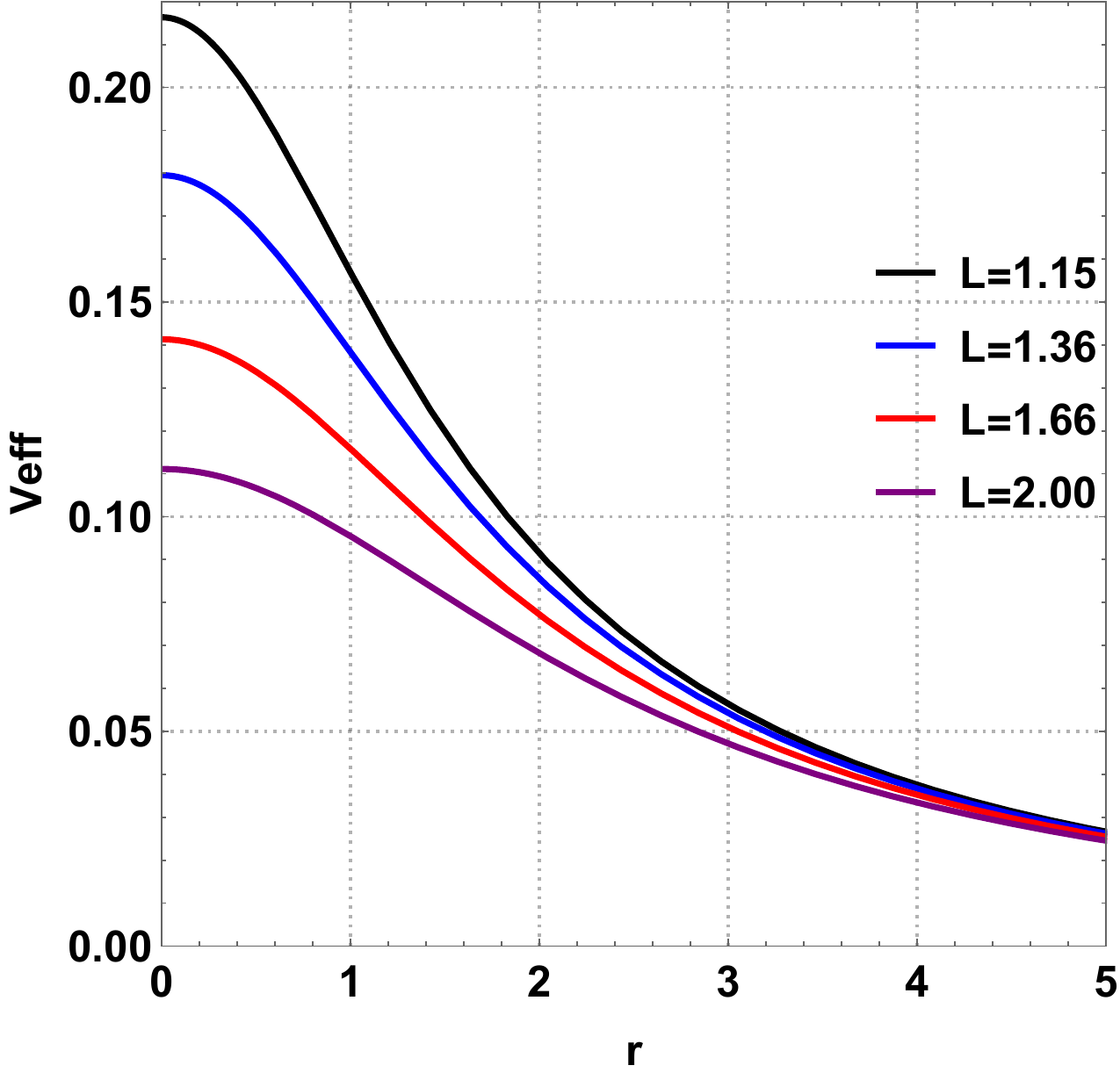}\label{fig3a}}
\hspace{1.2cm}
\subfigure[Lensing effect in modified null singularity spacetime.]
{\includegraphics[width=62mm]{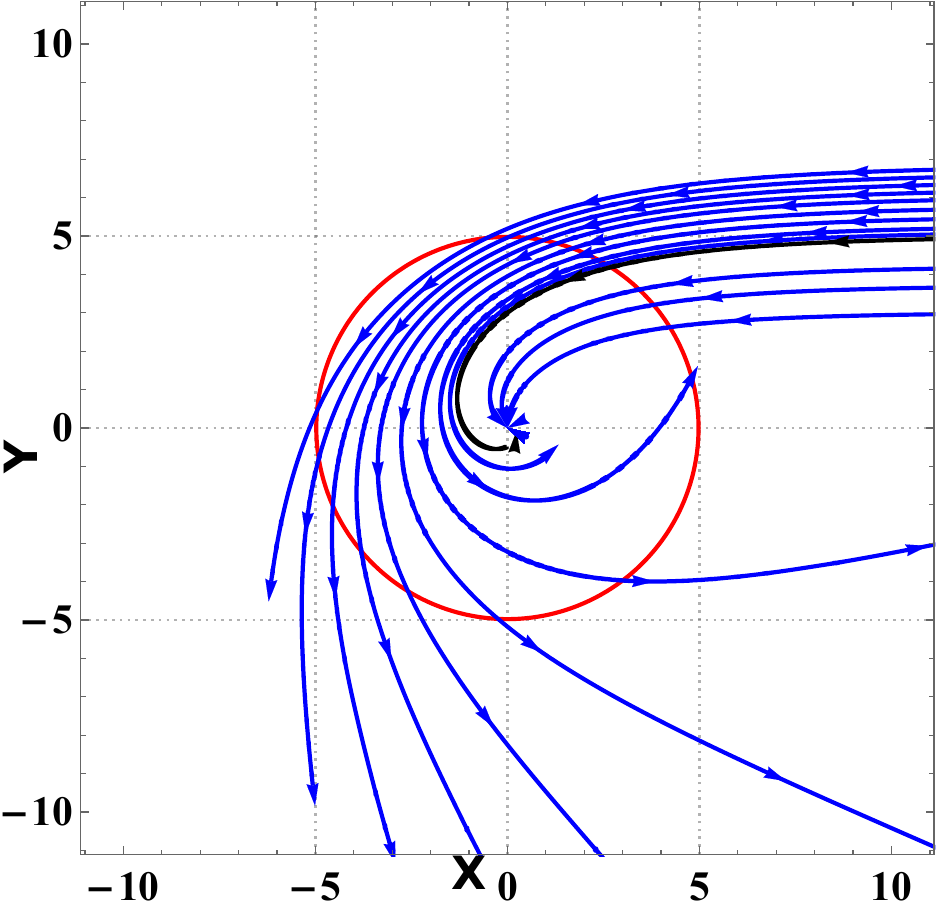}\label{fig3b}}\\
\hspace{0.2cm}
\subfigure[Intensity distribution in modified null singularity spacetime.]
{\includegraphics[width=68mm]{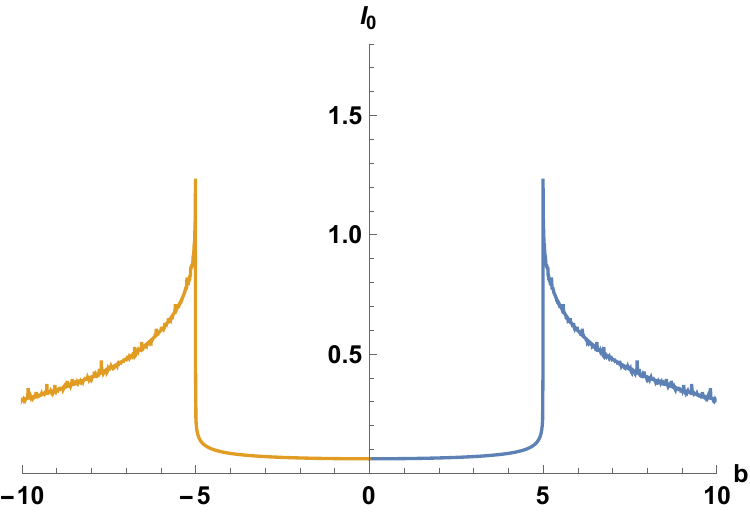}\label{fig3c}}
\hspace{0.2cm}
\subfigure[Shadow in modified null singularity spacetime.]
{\includegraphics[width=68mm]{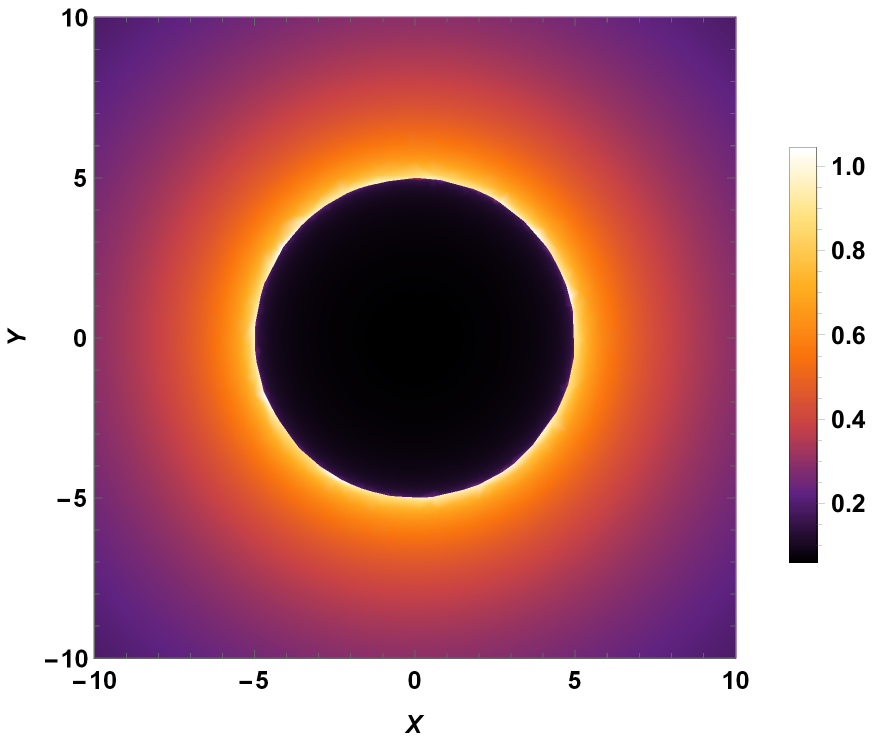}\label{fig3d}}
 \caption{In Figs.~(\ref{fig3a}), (\ref{fig3b}), (\ref{fig3c}), and (\ref{fig3d}) we show the nature of effective potentials of the null geodesics, light trajectory, intensity distribution and shadow image in modified null singularity spacetime. Effective potential of lightlike geodesics for different impact parameter in modified null singularity spacetime is shown in Fig.~(\ref{fig3a}). In Fig.~(\ref{fig3b}) the light trajectories in these spacetimes are shown, the blue lines are the null geodesics. In Figs.~(\ref{fig3c}) and (\ref{fig3d}), the intensity map in observer sky and the shadow of the central object are shown for the modified null singularity spacetime. The shadow shown in the right bottom corner is the shadow cast by the modified null singularity spacetime.}\label{fig3}
\end{figure*}

\begin{figure*}
\centering
\subfigure[The effective potential in modified charged null singularity spacetime.]
{\includegraphics[width=64mm]{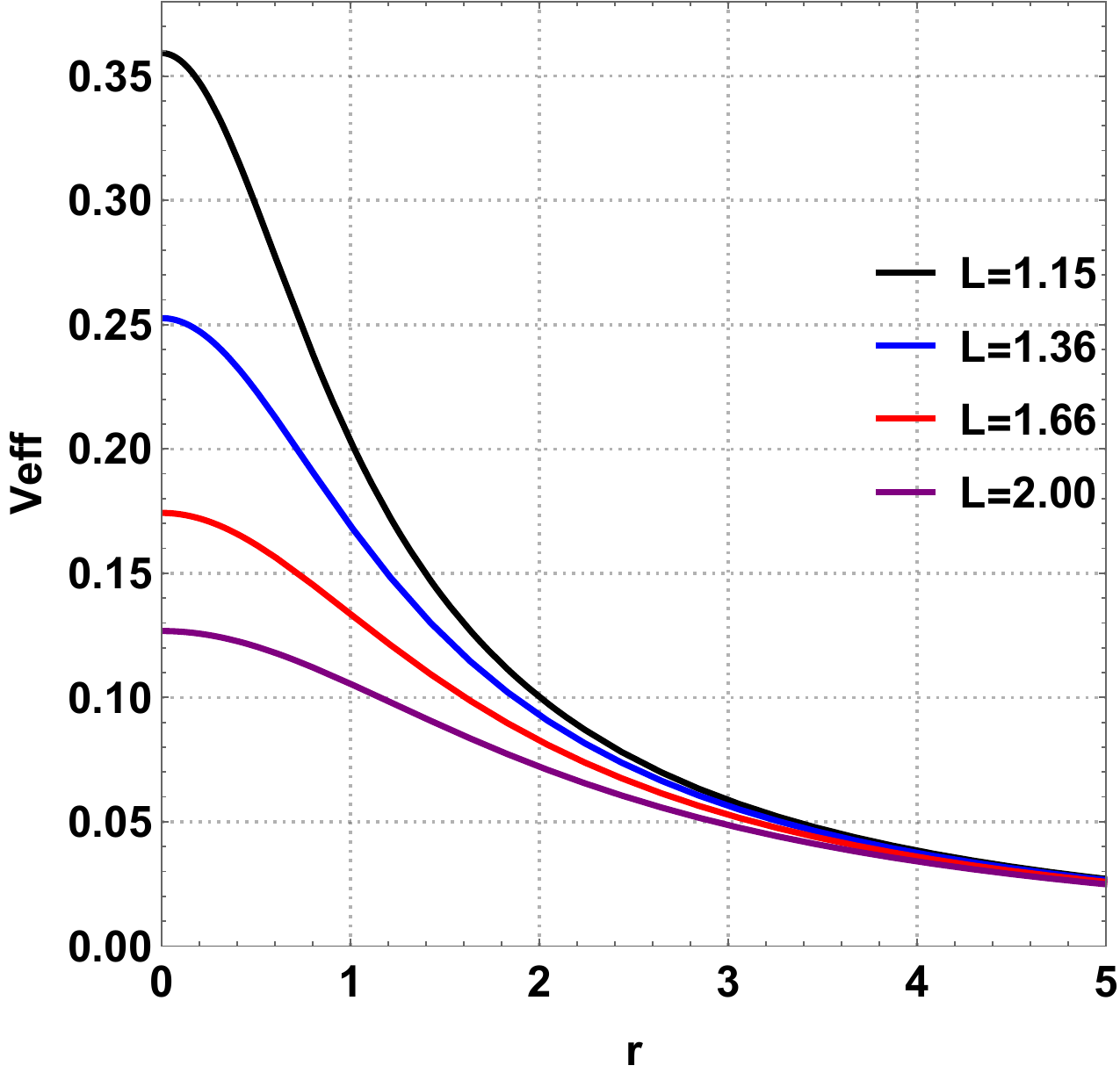}\label{fig4a}}
\hspace{1.2cm}
\subfigure[Lensing effect in modified charged null singularity spacetime.]
{\includegraphics[width=60mm]{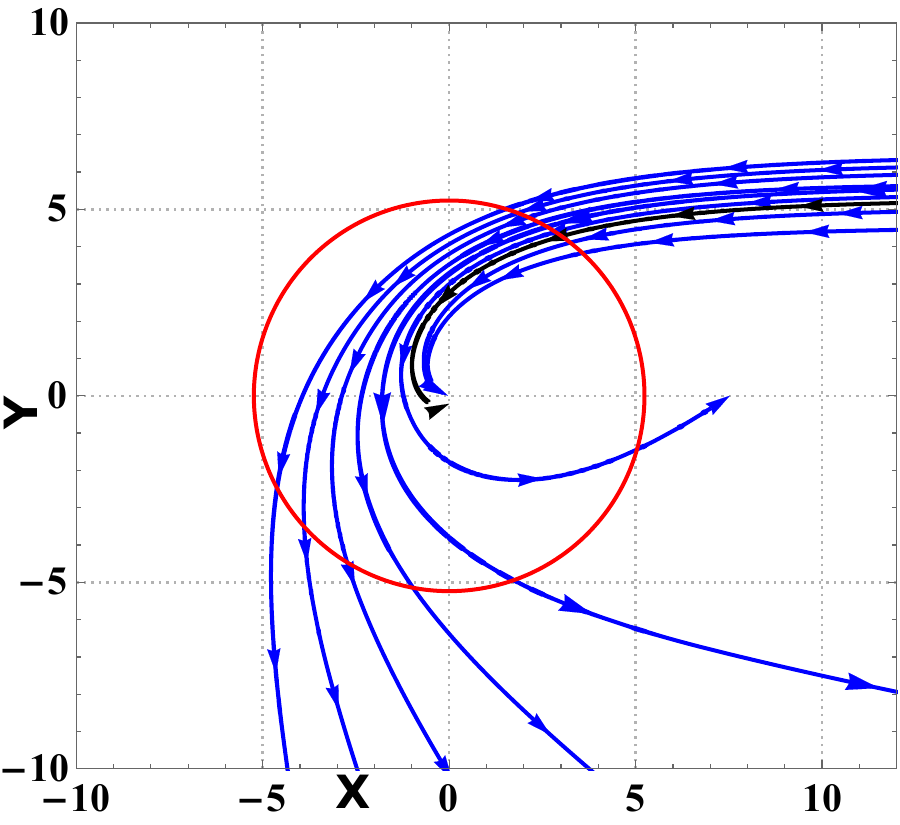}\label{fig4b}}\\
\hspace{0.2cm}
\subfigure[Intensity distribution in modified charged null singularity spacetime.]
{\includegraphics[width=82mm]{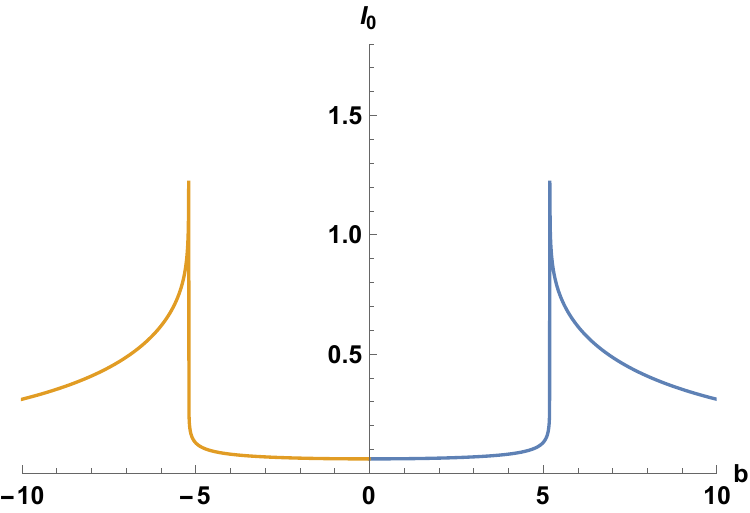}\label{fig4c}}
\hspace{0.2cm}
\subfigure[Shadow in modified charged null singularity spacetime.]
{\includegraphics[width=68mm]{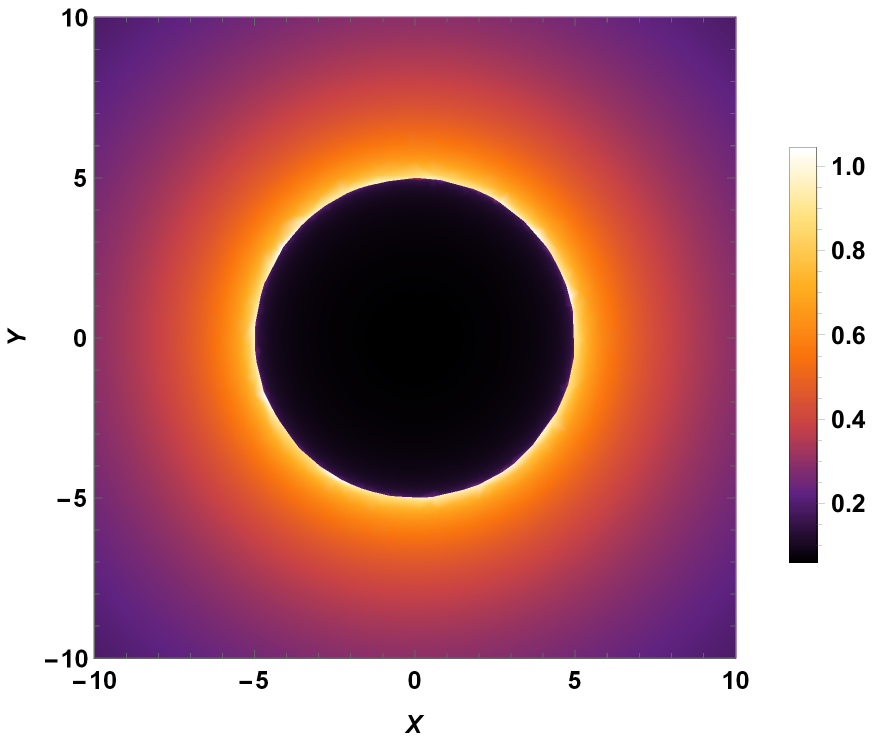}\label{fig4d}}
 \caption{In Figs.~(\ref{fig4a}), (\ref{fig4b}), (\ref{fig4c}), and (\ref{fig4d}) we show the nature of  effective potentials of the null geodesics, light trajectory, intensity distribution and shadow image in modified charged null singularity spacetime. Effective potential of lightlike geodesics for different impact parameter in modified charged null singularity spacetime is shown in Fig.~(\ref{fig4a}). In Fig.~(\ref{fig4b}) the light trajectories in these spacetimes are shown, the blue lines are the null geodesics. In Figs.~(\ref{fig4c}) and (\ref{fig4d}), the intensity map in observer sky and the shadow of the central object are shown for modified charged null singularity spacetime. The shadow shown in the right bottom corner is the shadow cast by the modified charged null singularity spacetime.}\label{fig4}
\end{figure*}

\section{Shadows in Regularized Spacetimes} \label{sec3}
Considering null geodesics in the equatorial plane (\( \theta = \pi/2 \)), the photon trajectory equation becomes \cite{Patel:2022vlu}:
\begin{equation}
    \frac{1}{b^2} = \frac{1}{h^2} \left( \frac{dr}{d\lambda} \right)^2 + V_{\text{eff}}\,,
\end{equation}
where \( V_{\text{eff}} = \frac{f(r)}{(r^2 +L^2)} \) is the effective potential, and \( b = \frac{h}{\gamma} \) is the impact parameter, defined in terms of the conserved energy \( \gamma \) and angular momentum \( h \). The null condition \( k^\mu k_\mu = 0 \), with \( k^\mu \) as the photon's four-momentum, is used here. We note that the effective potential possesses an extremum at the throat (r=0). Since different conventions exist in the literature regarding whether such throat orbits should be classified as exterior photon spheres (or light rings), we explicitly adopt the terminology used throughout this work and clarify the distinction between a conventional exterior photon sphere associated with unstable null circular orbits and the throat orbit appearing in the Simpson--Visser geometry. The implications of this distinction for the resulting image structure are discussed below.

The nature of the effective potential determines the circular photon orbits around the object. A local maximum \( V_{\text{eff}} \) corresponds to an unstable circular orbit; this defines the exterior photon sphere. The conditions for such an orbit at radius \( r_{\text{ph}} \) are:
\[
 b_{ph} = \frac{\sqrt{r_{ph}^2+L^2}}{\sqrt{f(r_{ph})}}\,, \quad V'_{\text{eff}}(r_{\text{ph}}) = 0, \quad V''_{\text{eff}}(r_{\text{ph}}) < 0\,.
\]
Similarly, a local minima of \( V_{\text{eff}} \) corresponds to a stable circular: this defines the anti-exterior photon sphere, \( r_{\text{aph}} \). The conditions for a stable orbit are similar to the case of a exterior photon sphere, but \( V''_{\text{eff}}(r_{\text{aph}})>0 \). If the effective potential has only a maximum, then \( b_{\text{ph}} \) gives the smallest impact parameter for which photons can escape. Light rays from distant sources with \( b < b_{\text{ph}} \) are captured and cannot reach the observer, resulting in a shadow; in the observer’s sky with a radius \( b_{\text{ph}} \). The value of the impact parameter at $r\to 0$ is $b_{0}$, is defined where \( V_{\text{eff}}(0) = \frac{1}{b^2} \), which leads to:
\begin{equation}
   b_{0}= b_{tp}(0) = \frac{L^2}{\sqrt{f(0)}}\,.
    \label{impact1}
\end{equation}
However, in spacetimes where no exterior photon sphere exists and where the effective potential diverges near the center, in such a case, shadows may not form in the same way. The effective potential remains finite at the origin, implies photons with energy \( E \geq V_{\text{eff}}(0) \) can reach the singularity. Thus, photons with \( b < b_{0} \) are trapped near the singularity, forming a shadow even in the absence of a exterior photon sphere. This shadow can be interpreted as the shadow cast by singularity itself. 
Orbit equation in terms of $u=1/r$ is given as follows:
\begin{equation}
    \frac{d^2 u}{d\phi^2} = \frac{2 L^2 u \psi}{b^2} - \xi u f(u) - \frac{\psi u^2}{2}\frac{df(u)}{du},
\end{equation}
where, $\psi = (1+ L^2 u^2)$ and $\xi = (1+2 L^2 u^2)$.
To properly visualize shadows in an astrophysical context, one must compute the observed intensity from surrounding accretion matter.
The position of the exterior photon sphere in the black bounce, and for $L<3M$, the exterior photon sphere is present in that spacetime. For $L<2M$, black bounce spacetime has an event horizon. In between $2M< L <3M$, a exterior photon sphere is present, but the event horizon is absent. Hence, in regular black bounce spacetime, without a black hole, a exterior photon sphere can exist. The behavior of lightlike geodesics around the black bounce spacetime when the exterior photon sphere is present is similar to the Schwarzschild spacetime, and when the exterior photon sphere is absent, the lightlike geodesics are identical to the null singularity spacetime as given in Fig. \ref{fig1b}. We have shown a plot for all figures in the coordinate radius and plot range is $r\geq0$. Because for $0\geq r$ it show the reflection symmetry. In Fig. \ref{fig1a}, the effective potential is shown for different $L$ values to illustrate the nature of potential change. The critical impact parameter at exterior photon sphere is $b_{ph}$. $b_{ph}$ is independent of $L$ values. Hence, shadow size remain constant irrespective of $L$. Therefore, for $L=3M$, $r_{ph}$, Fig. \ref{fig1c} shows intensity distribution versus impact parameter, and Fig. \ref{fig1d} shows shadow image for distant observer. \\

The position of the exterior photon sphere $(r_{ph})$ and antiexterior photon sphere $(r_{aph})$ are, $r_{ph} = M\sqrt{a+d}$ and $r_{aph} = M\sqrt{a-d}$, respectively, where, $a = 9/2 -x^2 - 2y^2$ and $d= (3/2)\sqrt{9-8y^2}$ \cite{Guo:2021wid}. Here, $x= L/M$ and $y = q/M$. The exterior photon sphere and the antiexterior photon sphere are present in spacetime when, 
\begin{equation}
    x \leq \sqrt{\frac{9}{2} - 2y^2 + \frac{3}{2}\sqrt{9-8y^2}},
\end{equation}
and 
\begin{equation}
    x \leq \sqrt{\frac{9}{2} - 2y^2 - \frac{3}{2}\sqrt{9-8y^2}},
\end{equation}
respectively. For real values of $x$, the range of $y$, $0<y \leq \frac{3}{2\sqrt{2}}$, which implies the $0<x \leq \frac{9}{4}$. Therefore, within the above given range, exterior photon sphere and anti-exterior photon sphere are present; above that range, a shadow is always formed without a exterior photon sphere. This range also indicates that, in the absence of an event and a Cauchy horizon, a exterior photon sphere and an anti-exterior photon sphere are present in the range $1< x < \frac{3}{2\sqrt{2}}$. At $x = \frac{3}{2\sqrt{2}}$, the extremal case where, $r_{ph} = r_{aph}$. However, for certain ranges, the anti-exterior photon sphere vanishes, but the exterior photon sphere is present in that spacetime,
\begin{equation}
    r_{ph} = 3M \sqrt[4]{1-\frac{8}{9}y^2}.\label{rphab}
\end{equation} Hence, from Eq. (\ref{rphab}), we can conclude that, in the absence of an antiexterior photon sphere, the exterior photon sphere is always smaller than $3M$. The critical impact parameter on the photon and anti-exterior photon spheres are,
\begin{equation}
    b_{ph} = \frac{(\frac{9}{2} -2y^2 + d)M}{\frac{9}{2} -y^2 + d -2 \sqrt{\frac{9}{2} -2y^2 + d}},
\end{equation}
and
\begin{equation}
    b_{ph} = \frac{(\frac{9}{2} -2y^2 - d)M}{\frac{9}{2} -y^2 - d -2 \sqrt{\frac{9}{2} -2y^2 - d}},
\end{equation}
respectively. From equations of critical impact parameter, we conclude that the size of the shadow does not change over $L$ values. In certain cases, the local maxima of potential $V_{eff}(0)>V_{eff}(r_{ph})$, which indicates that the shadow cast by the singularity itself, instead of the exterior photon sphere. Impact parameter values at $r=0$ is given as follows:
\begin{equation}
    b_{0} = \frac{x^2 M}{\sqrt{x^2-2x+y^2}}.
\end{equation}
Here, because of the exterior photon sphere a photon ring forms in the observer's sky. However, the size of the shadow is smaller than the photon ring formed by the exterior photon sphere. For a critical case that is for $V_{eff}(0)=V_{eff}(r_{ph})$, $b_{0}=b_{ph}$. This provides a constraint on the $x$ that is,
\begin{equation}
    x = \sqrt{\frac{9}{2}-2y^2 + a}.
\end{equation}
When $V_{eff}(0)<V_{eff}(r_{ph})$, the exterior photon sphere plays a crucial role in the formation of the shadow and its size. The effective potential and the behaviour of lightlike geodesics in charged black bounce spacetime are shown in Fig.~(\ref{fig2a}) and Fig.~(\ref{fig2b}) respectively, where we take mass $M=1$. The red circle in Fig.~(\ref{fig1b}), Fig.~(\ref{fig2b}), Fig.~(\ref{fig3b}) and Fig.~(\ref{fig4b}) shows the shadow size ($b_{ph}$) in given spacetime configuration. The black and blue lines show the light ray for the critical impact parameter and lensed light rays. 

In a modified null singularity spacetime, the analysis of the position of the exterior photon sphere provides $r_{ph}=0$. Hence, the critical impact parameter is $b_{0}=(1+x)M$. Therefore, the size of the shadow depends on the value of $x$. This provides freedom for providing observational range on the value of $x$. Fig.\,\ref{fig3a} shows the potential of lightlike geodesics varying with $L$ values. For all values of $L$ exterior photon sphere is absent. 
Fig.\,\ref{fig3b} shows the lensing of null geodesics, which is similar to that of the null-singularity spacetime. For the specific parameter choice $L=4.2$, the corresponding shadow radius is approximately $3\sqrt{3}M$ as illustrated by the intensity profile in Fig.\,\ref{fig3c} and the shadow image in Fig.\,\ref{fig3d}.

In a modified charged null singularity spacetime, the analysis of the position of the exterior photon sphere provides $r_{ph}=0$. Hence, the critical impact parameter is,
\begin{equation}
    b_{0} = \frac{x^2(1+x)M}{\sqrt{x^4+(1+x)^2y^2}}.
\end{equation}
Therefore, the size of the shadow depends on the value of $x$ and $y$. In particular, for fixed $x$, Eq.~(22) predicts that the shadow radius decreases monotonically with increasing $|y|$. This provides freedom for providing observational range on the value of $x$ and $y$. Fig.\,\ref{fig4a} shows the potential of lightlike geodesics varying with $L$ and $q$ values. For all values of $L$ and $q$ exterior photon sphere is absent. Fig.\,\ref{fig4b} shows the lensing of null geodesics, which is similar to that of the null-singularity spacetime. For the specific parameter choice $L=4.2$ and $Q=0.13$ using Eq.(22), the corresponding shadow radius is approximately $3\sqrt{3}M$ as illustrated by the intensity profile in Fig.\,\ref{fig4c} and the shadow image in Fig.\,\ref{fig4d}.

\begin{figure*}
\centering
\subfigure[]
{\includegraphics[width=8cm]{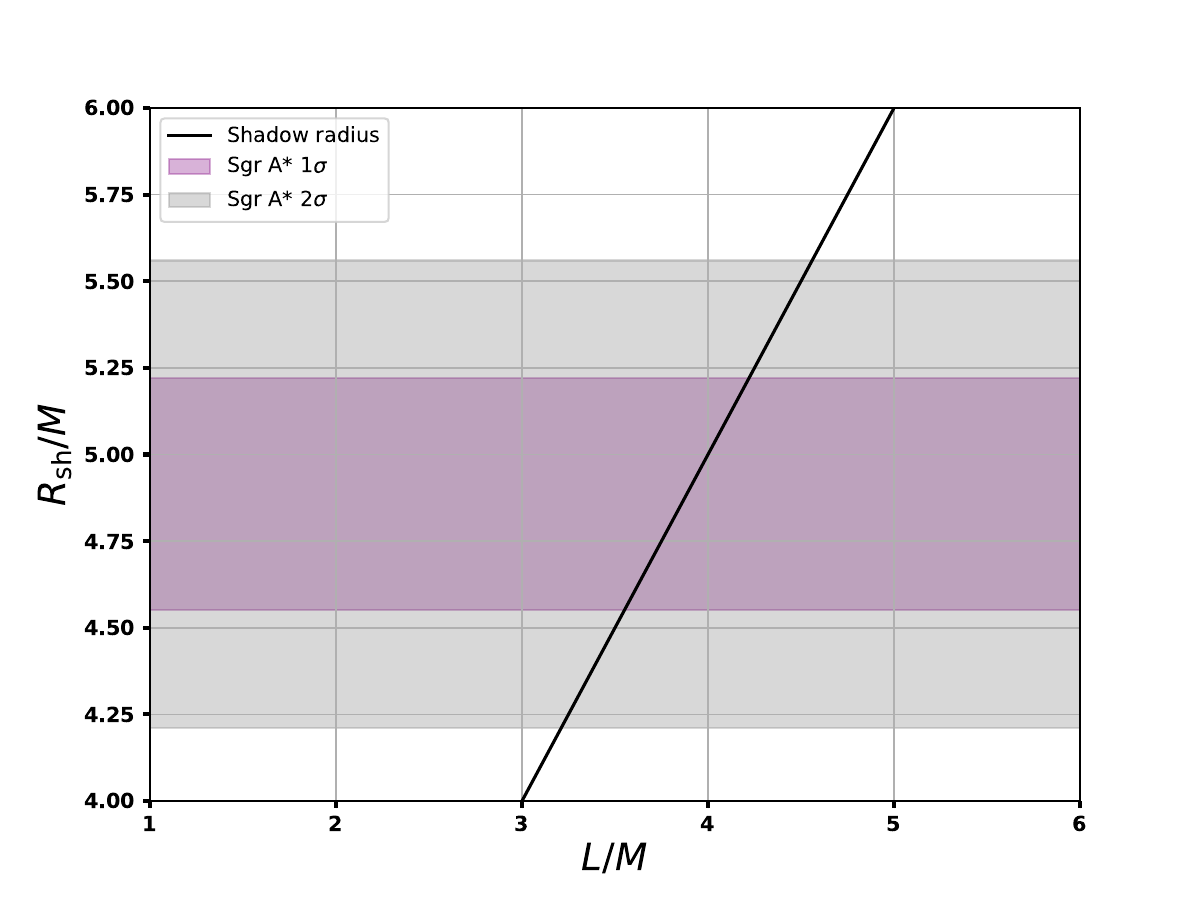}\label{a}}
\hspace{0.2cm}
\subfigure[]
{\includegraphics[width=8cm]{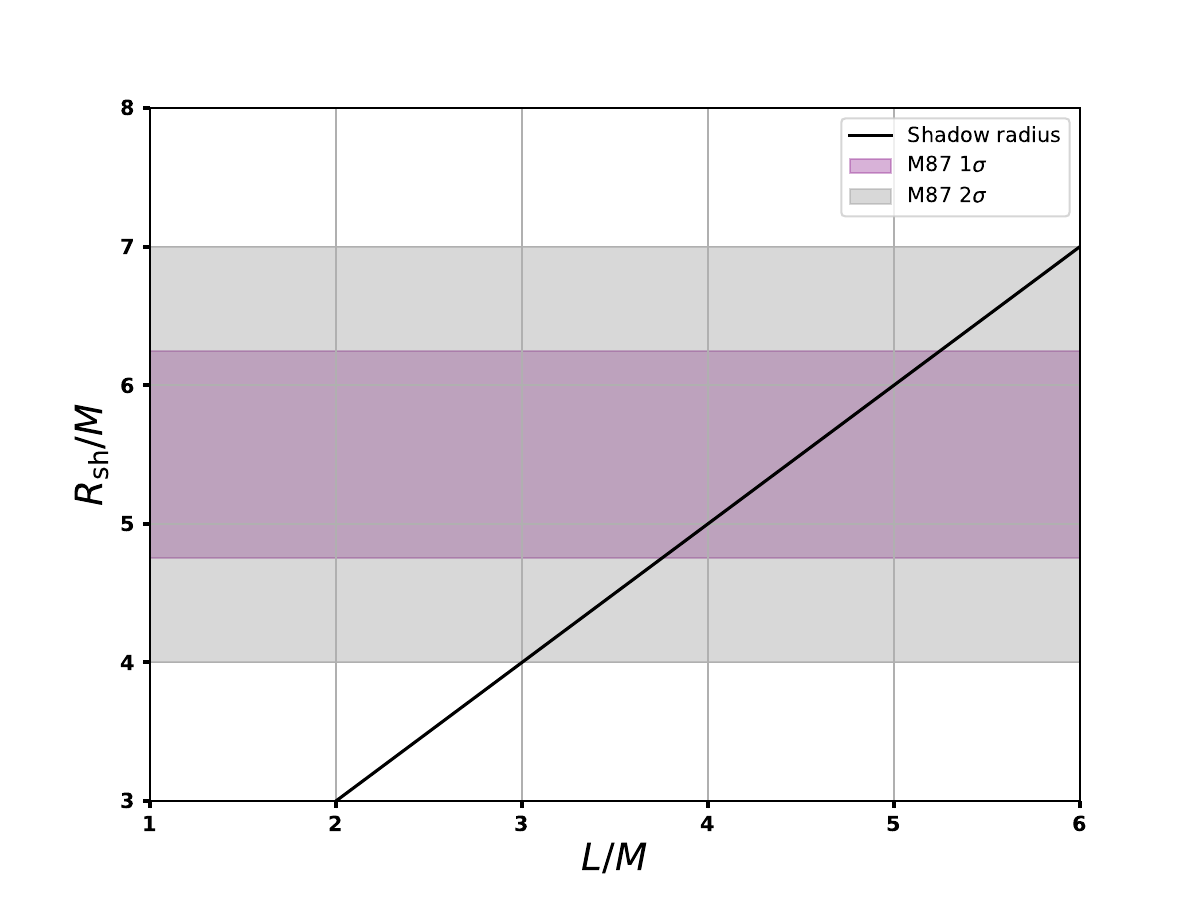}\label{b}}
\hspace{0.2cm}
\subfigure[]
{\includegraphics[width=8cm]{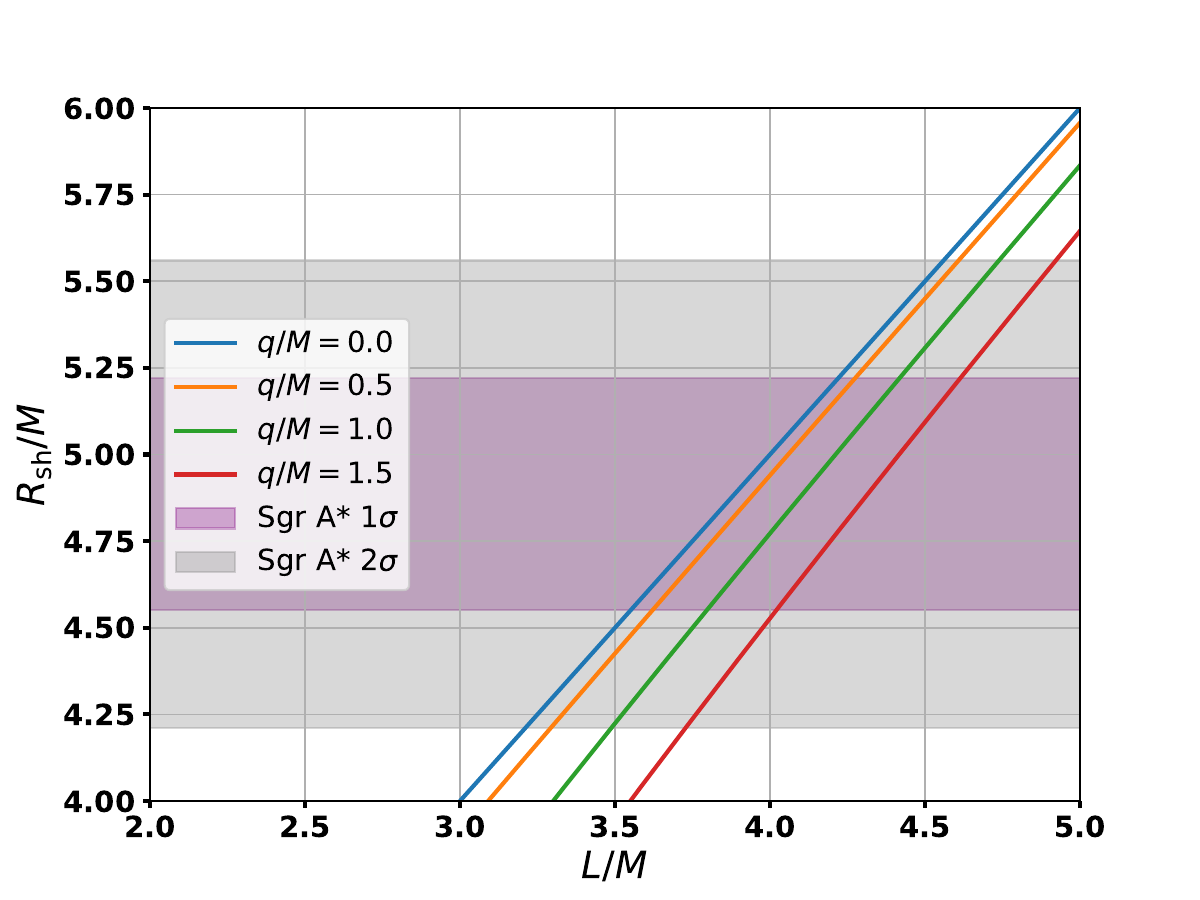}\label{c}}
\hspace{0.2cm}
\subfigure[]
{\includegraphics[width=8cm]{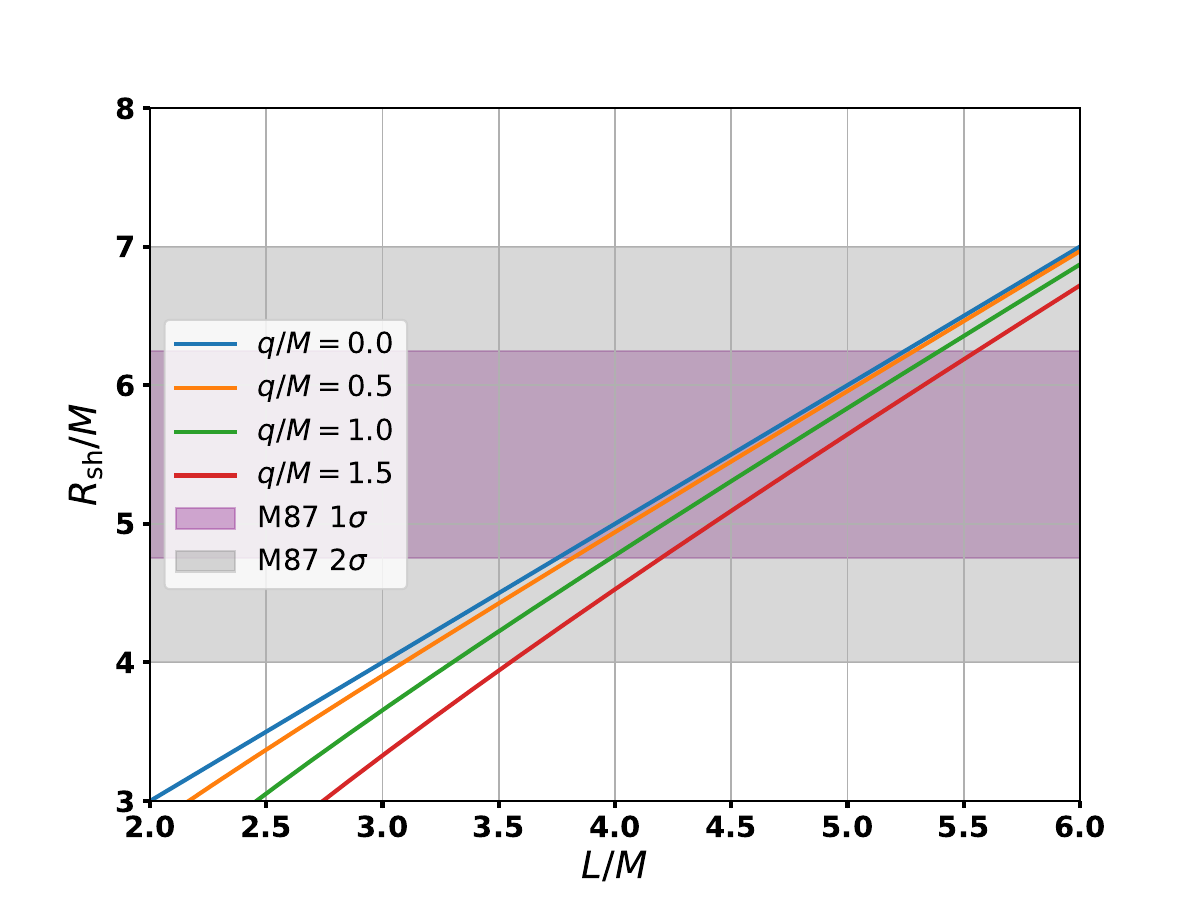}\label{d}}
 \caption{Shadow radius in null regular geometry compared with Sgr $A^{*}$ and M87 for $1\sigma$ and $2\sigma$ range derived from the EHT observations are presented in Figs.\,\ref{a} and \ref{b} respectively. Shadow radius in charged null regular geometry compared with Sgr $A^{*}$ and M87 for $1\sigma$ and $2\sigma$ range derived from the EHT observations are presented in Figs.\,\ref{c} and \ref{d} respectively.}\label{observation}
\end{figure*}

\section{Intensity distributions around compact objects}\label{sec4}
In the present analysis, the simplified emission model is performed only on one asymptotically flat side of the regularized spacetime. The specific intensity is integrated from the throat ($r=0$) to the distant observer, and no emission from the opposite asymptotic region is considered. Consequently, the resulting central dark region is model dependent and reflects the adopted emissivity and boundary conditions rather than photon capture by an event horizon. In this work, we assume spherically symmetric, radially infalling thin accretion matter that emits monochromatic radiation. The emissivity (in the emitter’s frame) is modeled as:
\begin{equation}
    j(\nu_e) \propto \frac{\delta(\nu_e - \nu_*)}{(r^2 +L^2)}\,,
\end{equation}
where \( \nu_e \) is the photon frequency in the emitter's rest frame. The observed intensity on the distant sky, in terms of image-plane coordinates \( (X, Y) \), is given by \cite{bambi}:
\begin{equation}
    I_{\nu_o}(X,Y) = \int_{\gamma} g^3 j(\nu_e) \, dl_{\text{prop}}\,,
    \label{eq31}
\end{equation}
with redshift factor \( g = \nu_o / \nu_e \), and proper length element in the emitter’s frame:
\[
dl_{\text{prop}} = -k_\alpha u^\alpha_e d\lambda\,.
\]

Here, \( k^\mu \) is the photon’s four-momentum, \( u^\mu_e \) is the four-velocity of the emitter, and \( \lambda \) is the affine parameter. For a distant static observer, \( u^\mu_o = (1, 0, 0, 0) \), the redshift factor becomes:
\begin{equation}
    g = \frac{k_\alpha u^\alpha_o}{k_\beta u^\beta_e}\,.
\end{equation}

For radial infall, the emitter's four-velocity components are:
\begin{equation}
    u^t_e = \frac{1}{g_{tt}}, \quad
    u^r_e = -\sqrt{\frac{1 - g_{tt}}{g_{tt} g_{rr}}}, \quad
    u^\theta_e = u^\phi_e = 0\,.
\end{equation}
Substituting these gives the redshift factor as:
\begin{equation}
    g = \frac{1}{\frac{1}{g_{tt}} - \frac{k_r}{k_t} \sqrt{\frac{1 - g_{tt}}{g_{tt} g_{rr}}}}\,,
\end{equation}
where
\begin{equation}
    \frac{k^r}{k^t} = \sqrt{\frac{g_{tt}}{g_{rr}} \left( 1 - \frac{g_{tt} b^2}{r^2 + L^2} \right)}\,.
\end{equation}

The final expression for the observed intensity becomes \cite{bambi}:
\begin{equation}
    I_o(X,Y) \propto -\int_{\gamma} \frac{g^3 k_t \, dr}{(r^2 + L^2) k^r}\,,
    \label{intensity}
\end{equation}
where \( X^2 + Y^2 = b^2 \). Using Eq.~\eqref{intensity}, the shadow profile as seen by a distant observer can be simulated.
\subsection{Phenomenological Comparison with EHT Angular Scales}
In this section, we perform a phenomenological comparison between the predicted shadow sizes and the angular scales reported by the EHT observations. Our objective is to examine the observational consistency of the model rather than to perform a statistical parameter estimation or derive quantitative confidence intervals.
When it comes to observation, the BH shadow diameter is important since high-resolution very long baseline interferometry (VLBI) observations, particularly those carried out by the EHT Collaboration, can directly constrain it. The EHT has produced unprecedented measurements of the shadow size by imaging the vicinity of supermassive compact objects like Sgr A* and M87*. Based on these observational data, we can do the phenomenological Comparison of the parameters $M, L$, and $q$. These data serve as a crucial testbed for examining departures from the Kerr metric's predictions. They are among the most direct probes of the strong-gravity regime near the event horizon. The appropriate limitations on the dimensionless shadow radius $ rsh/M $ can be inferred as follows using the observed angular diameters and the mass estimates of these BHs \cite{Vagnozzi:2022moj,Kala:2025fld}:
\begin{equation*}
\begin{aligned}
\text{Sgr A}^*:\quad
&\begin{cases}
1\sigma: & 4.55 \lesssim \dfrac{r_{\rm sh}}{M} \lesssim 5.22, \\
2\sigma: & 4.21 \lesssim \dfrac{r_{\rm sh}}{M} \lesssim 5.56.
\end{cases}
\\[1em] 
\text{M87}^*:\quad
&\begin{cases}
1\sigma: & 4.75 \lesssim \dfrac{r_{\rm sh}}{M} \lesssim 6.25, \\
2\sigma: & 4.00 \lesssim \dfrac{r_{\rm sh}}{M} \lesssim 7.00.
\end{cases}
\end{aligned}
\end{equation*}

In Fig\,(\ref{observation}), Shadow radius in the null and charged null regularized spacetime compared with the EHT observational bounds for Sgr $A^{*}$ and M87. Figs.\,\ref{a} and \ref{b} show the variation of the normalized shadow radius $Rsh/M$ with respect to the parameter $L/M$ for the neutral case, while the shaded regions represent the $1\sigma$ and $2\sigma$ observational range derived from the EHT measurements of Sgr $A^{*}$ and M87, respectively. Figs.\,\ref{c} and \ref{d} correspond to the charged null regular geometry, where the influence of the charge parameter $q/M$ on the shadow radius is demonstrated for different values $q/M = 0, 0.5, 1.0,$ and $1.5$. The shadow radius decreases slightly with increasing charge, indicating a modest deviation from the Schwarzschild limit. The colored bands denote the observational ranges of the shadow angular diameter reported by the EHT Collaboration, within which the theoretical shadow size is compatible with the measured angular scales. Overall, the comparison shows that both Sgr $A^{*}$ and M87 phenomenological comparison allowed parameters space of the null regular geometry, with the neutral case lying well within the observational limits and higher charges leading to marginally larger deviations.

\section{Conclusion}
\label{result}
    In this work, we examined the behavior of light rays and the resulting shadow structure in a class of regularized spacetimes constructed through the Simpson–Visser procedure. Since strong-gravity lensing carries crucial information about the nature of the central object, we analyzed how these geometries produce shadow-like central intensity depressions (hereafter referred to as shadow-like images) despite lacking conventional exterior photon spheres. Our results show that the presence of a bright photon ring does not necessarily imply that the shadow boundary is determined by an exterior photon sphere. We identify configurations in which both an exterior photon sphere and an anti-exterior photon sphere are present, yet the shadow-like central dark region is governed primarily by the regularized core geometry. In these cases, although the bright photon ring is associated with the exterior photon sphere, the shadow-like dark region remains significantly smaller than the ring. This demonstrates that the edge of the observed dark region does not necessarily provide direct evidence for the existence or the location of an exterior photon sphere.\\

    A key outcome of our analysis is that the simulated shadow-like images produced by these modified null singularity spacetimes closely resemble those of the Schwarzschild black hole, even though their causal structures differ fundamentally and no exterior photon sphere is required to form a shadow. We further demonstrate that the location of the exterior photon sphere, when present, does not uniquely determine the shadow size. This may have implications for interpreting high-resolution observations of compact objects, including Sgr A*, where distinguishing between classical black holes and alternative compact objects remains a central challenge.\\

    Several studies suggest that the regularization parameter L may give quantum gravitational information. In this interpretation, L smooths the classical singularity by introducing effects that become dominant near the gravitational center. If such quantum corrections influence photon trajectories in the strong-field region, they may leave observable imprints on the shadow cast by the regularized core. Identifying and characterizing these properties would require a more detailed investigation of the precise features that lie within the shadow structure.\\

    We emphasize that the present calculations are based on an idealized optically thin emission model intended to investigate the qualitative optical signatures of the considered spacetimes. A quantitative comparison with observations requires metric-specific plasma dynamics, general-relativistic radiative-transfer calculations, and GRMHD simulations together with comparisons to observational likelihoods, which remain beyond the scope of the present work.\\

    Overall, our study indicates that shadow-like image structures comparable in size to those inferred from the EHT observations can, in principle, arise in spacetimes without event horizons or exterior photon spheres under the adopted optically thin emissivity model. Similar observational features can emerge from exotic compact configurations, including regularized null singularities. This is possibly linked to mechanisms of singularity resolution in quantum gravity. These results demonstrate that the predicted shadow angular diameters of the proposed geometries are compatible with the angular-scale measurements reported by the EHT Collaboration. However, a comprehensive assessment of their observational viability requires additional observables, such as brightness distributions, higher-order lensed-ring structures, polarization, and near-infrared flux measurements, which are beyond the scope of the present work.\\

 \section{ACKNOWLEDGMENTS} 
The authors thank the anonymous referee for their valuable comments and suggestions. VP acknowledges the support of the Council of Scientific and Industrial Research (CSIR, India; Ref: 09/1294(18267)/2024-EMR-I) for financial support. 

\section*{Appendix A: Numerical procedure for ray tracing and image construction}

In this appendix, we summarize the numerical procedure employed to generate the shadow images presented in this work. The implementation is general and is applicable to any static, spherically symmetric spacetime described by the metric
\begin{equation}
ds^{2}=-f(r)dt^{2}+\frac{dr^{2}}{f(r)}+R^{2}(r)\left(d\theta^{2}+\sin^{2}\theta d\phi^{2}\right),
\end{equation}
where the metric function $f(r)$ and the areal radius $R(r)$ are specified for each spacetime considered.

For null geodesics, the conserved energy $\gamma$ and angular momentum ($L_{z}$) define the impact parameter
\begin{equation}
b=\frac{L_{z}}{\gamma}.
\end{equation}

The radial motion satisfies
\begin{equation}
\left(\frac{dr}{d\lambda}\right)^{2} = \gamma^{2} - V_{\rm eff}(r),
\end{equation}
where (V$_{eff}$(r)) denotes the effective potential corresponding to the chosen spacetime.

For a distant observer located at radius ($r_{obs}$), each pixel on the image plane is assigned Cartesian coordinates $(X,Y)$. These are converted into the corresponding impact parameter,
\begin{equation}
b=\sqrt{X^{2}+Y^{2}},
\end{equation}
and the initial conditions of the photon trajectory are determined accordingly.

For every image-plane pixel, the null geodesic equations are integrated numerically using a fourth-order Runge--Kutta algorithm with adaptive step size. The integration proceeds backward from the observer toward the compact object until one of the following stopping conditions is satisfied:

\begin{enumerate}
\item the photon reaches the emitting region, turning point of null geodesics 
\item the photon crosses the event horizon (if present),
\item the photon passes through the throat or reaches the opposite asymptotic region (for wormhole geometries),
\item or the integration exceeds the prescribed numerical tolerance.
\end{enumerate}

The observed intensity is evaluated using the standard optically thin radiative transfer approximation,

\begin{equation}
I_{\rm obs} = \int g^{3} j(\nu_{\rm e})dl_{\rm prop},
\end{equation}

where (g=$\nu_{obs}$/$\nu_{e}$) is the redshift factor, (j($\nu_{e}$)) denotes the emissivity in the emitter frame, and (d$l_{prop}$) is the infinitesimal proper length measured in the rest frame of the emitting gas. In the present work, we adopt the monochromatic, optically thin emission model with a radially falling velocity profile described in Sec.~IV.

The image is constructed by repeating this procedure for all pixels of a uniformly spaced Cartesian grid covering the observer's image plane. The resulting intensity values are normalized by the maximum intensity within each image and are displayed using the color scale shown in the figures.

Unless otherwise stated, numerical calculations employ geometrized units ((G=c = M=1)). The observer is placed sufficiently far from the compact object (($r_{obs}\gg M$)) so that asymptotic flatness provides an excellent approximation, and the integration domain is chosen large enough that further increases produce no visible change in the resulting images. Convergence was verified by repeating the calculations with finer image-plane resolutions and smaller integration step sizes, yielding indistinguishable shadow boundaries and intensity distributions.


\begin{thebibliography}{99}
\bibitem{EventHorizonTelescope:2019dse}
K.~Akiyama \textit{et al.} [Event Horizon Telescope],
\href{https://doi.org/10.3847/2041-8213/ab0ec7}{Astrophys. J. Lett. \textbf{875}, L1 (2019)}

\bibitem{EventHorizonTelescope:2019uob}
K.~Akiyama \textit{et al.} [Event Horizon Telescope],
\href{https://doi.org/10.3847/2041-8213/ab0c96}{Astrophys. J. Lett. \textbf{875}, no.1, L2 (2019)}

\bibitem{EventHorizonTelescope:2019jan}
K.~Akiyama \textit{et al.} [Event Horizon Telescope],
\href{https://doi.org/10.3847/2041-8213/ab0c57}{Astrophys. J. Lett. \textbf{875}, no.1, L3 (2019)}

\bibitem{EventHorizonTelescope:2019ths}
K.~Akiyama \textit{et al.} [Event Horizon Telescope],
\href{https://doi.org/10.3847/2041-8213/ab0e85}{Astrophys. J. Lett. \textbf{875}, no.1, L4 (2019)}

\bibitem{EventHorizonTelescope:2019pgp}
K.~Akiyama \textit{et al.} [Event Horizon Telescope],
\href{https://doi.org/10.3847/2041-8213/ab0f43}{Astrophys. J. Lett. \textbf{875}, no.1, L5 (2019)}

\bibitem{EventHorizonTelescope:2019ggy}
K.~Akiyama \textit{et al.} [Event Horizon Telescope],
\href{https://doi.org/10.3847/2041-8213/ab1141}{Astrophys. J. Lett. \textbf{875}, no.1, L6 (2019)}

\bibitem{EventHorizonTelescope:2020qrl}
D.~Psaltis \textit{et al.} [Event Horizon Telescope],
\href{https://journals.aps.org/prl/abstract/10.1103/PhysRevLett.125.141104}{Phys. Rev. Lett. \textbf{125}, no.14, 141104 (2020).}

\bibitem{EventHorizonTelescope:2022xnr}
K.~Akiyama \textit{et al.} [Event Horizon Telescope],
\href{doi:10.3847/2041-8213/ac6674}{Astrophys. J. Lett. \textbf{930}  no.2, L12 (2022).}

\bibitem{EventHorizonTelescope:2022vjs}
K.~Akiyama \textit{et al.} [Event Horizon Telescope],
\href{doi:10.3847/2041-8213/ac6675}{Astrophys. J. Lett. \textbf{930}  no.2, L13 (2022).}

\bibitem{EventHorizonTelescope:2022wok}
K.~Akiyama \textit{et al.} [Event Horizon Telescope],
\href{doi:10.3847/2041-8213/ac6429}{Astrophys. J. Lett. \textbf{930}  no.2, L14 (2022).}

\bibitem{EventHorizonTelescope:2022exc}
K.~Akiyama \textit{et al.} [Event Horizon Telescope],
\href{doi:10.3847/2041-8213/ac6736}{Astrophys. J. Lett. \textbf{930}  no.2, L15 (2022).}

\bibitem{EventHorizonTelescope:2022urf}
K.~Akiyama \textit{et al.} [Event Horizon Telescope],
\href{doi:10.3847/2041-8213/ac6672}{Astrophys. J. Lett. \textbf{930} no.2, L16 (2022).}

\bibitem{EventHorizonTelescope:2022xqj}
K.~Akiyama \textit{et al.} [Event Horizon Telescope],
\href{doi:10.3847/2041-8213/ac6756}{Astrophys. J. Lett. \textbf{930}  no.2, L17 (2022).}

\bibitem{EventHorizonTelescope:2022gsd}
J.~Farah \textit{et al.} [Event Horizon Telescope],
\href{doi:10.3847/2041-8213/ac6615}{Astrophys. J. Lett. \textbf{930}  no.2, L18 (2022).}

\bibitem{EventHorizonTelescope:2022ago}
M.~Wielgus \textit{et al.} [Event Horizon Telescope],
\href{doi:10.3847/2041-8213/ac6428}{Astrophys. J. Lett. \textbf{930}  no.2, L19 (2022).}

\bibitem{EventHorizonTelescope:2022tzy}
A.~E.~Broderick \textit{et al.} [Event Horizon Telescope],
\href{doi:10.3847/2041-8213/ac6584}{Astrophys. J. Lett. \textbf{930} no.2, L21 (2022).}

\bibitem{Eichhorn:2022oma}
A.~Eichhorn, A.~Held and P.~V.~Johannsen,
\href{https://iopscience.iop.org/article/10.1088/1475-7516/2023/01/043}{JCAP \textbf{01}, 043 (2023).}

\bibitem{Abdikamalov:2019ztb} 
A.~B.~Abdikamalov, A.~A.~Abdujabbarov, D.~Ayzenberg, D.~Malafarina, C.~Bambi and B.~Ahmedov,
\href{https://journals.aps.org/prd/abstract/10.1103/PhysRevD.100.024014}{Phys.\ Rev.\ D {\bf 100}, no. 2, 024014 (2019).}

\bibitem{Broderick:2024vjp}
A.~E.~Broderick and K.~Salehi,
\href{https://iopscience.iop.org/article/10.3847/1538-4357/ad90aa}{Astrophys. J. \textbf{977}, no.2, 249 (2024).}

\bibitem{atamurotov_2015}
F. Atamurotov, B. Ahmedov, and A. Abdujabbarov, 
\href{https://journals.aps.org/prd/abstract/10.1103/PhysRevD.92.084005}{Phys. Rev. D {\bf 92}, 084005 (2015).}

\bibitem{Vagnozzi:2019apd} 
S.~Vagnozzi and L.~Visinelli,
\href{https://journals.aps.org/prd/abstract/10.1103/PhysRevD.100.024020}{Phys.\ Rev.\ D {\bf 100}, no. 2, 024020 (2019).}

\bibitem{Gyulchev:2019tvk} 
  G.~Gyulchev, P.~Nedkova, T.~Vetsov and S.~Yazadjiev,
\href{https://journals.aps.org/prd/abstract/10.1103/PhysRevD.100.024055}{Phys.\ Rev.\ D {\bf 100}, no. 2, 024055 (2019).}

\bibitem{Narayan} 
  Ramesh Narayan, Michael D. Johnson, and Charles F. Gammie,
\href{https://iopscience.iop.org/article/10.3847/2041-8213/ab518c}{The Astrophysical Journal Letters, Volume 885, Number 2.} 

\bibitem{Gralla:2019xty} 
S.~E.~Gralla, D.~E.~Holz and R.~M.~Wald,
\href{https://journals.aps.org/prd/abstract/10.1103/PhysRevD.100.024018}{Phys.\ Rev.\ D {\bf 100}, no. 2, 024018 (2019).}

\bibitem{abdujabbarov_2015b} 
A. A. Abdujabbarov, L. Rezzolla, and B. J. Ahmedov, 
\href{https://academic.oup.com/mnras/article/454/3/2423/1196174}{Mon. Not. R. Astron. Soc. {\bf 454}, 2423 (2015).}

\bibitem{Li:2021}
G.~P.~Li and K.~J.~He,
\href{https://iopscience.iop.org/article/10.1088/1475-7516/2021/06/037}{JCAP \textbf{06}, 037 (2021).}

\bibitem{KumarWalia:2024omf}
R.~Kumar Walia, P.~Kocherlakota, D.~O.~Chang and K.~Salehi,
\href{https://journals.aps.org/prd/abstract/10.1103/PhysRevD.111.104074}{Phys. Rev. D \textbf{111}, no.10, 104074 (2025).}

\bibitem{Salehi:2024cim}
K.~Salehi, R.~Kumar Walia, D.~O.~Chang and P.~Kocherlakota,
\href{https://journals.aps.org/prd/abstract/10.1103/PhysRevD.111.104057}{Phys. Rev. D \textbf{111}, no.10, 104057 (2025).}

\bibitem{Ovgun:2025stp}
A.~{\"O}vg{\"u}n and M.~Fathi,
\href{https://www.sciencedirect.com/science/article/pii/S055032132500272X?via%3Dihub}{Nucl. Phys. B \textbf{1018}, 117063 (2025).}

\bibitem{Pedrotti:2025idg}
D.~Pedrotti and M.~Calz{\`a},
\href{https://journals.aps.org/prd/abstract/10.1103/1q35-mjjz}{Phys. Rev. D \textbf{111}, no.12, 12 (2025).}

\bibitem{Nalui:2025wiw}
S.~Nalui and S.~Bhattacharya,
\href{https://www.sciencedirect.com/science/article/pii/S0370269325000218?via%3Dihub}{Phys. Lett. B \textbf{861}, 139261 (2025).}

\bibitem{HassanPuttasiddappa:2025tji}
P.~Hassan Puttasiddappa, D.~C.~Rodrigues and D.~F.~Mota,
\href{https://link.springer.com/article/10.1140/epjc/s10052-025-14721-w}{Eur. Phys. J. C \textbf{85}, no.9, 974 (2025).}

\bibitem{Pantig:2024lpg}
R.~C.~Pantig, S.~Kala, A.~{\"O}vg{\"u}n and N.~J.~L.~S.~Lobos,
\href{https://www.worldscientific.com/doi/10.1142/S0219887825502408}{ https://doi.org/10.1142/S0219887825502408.}

\bibitem{Bambi:2019tjh}
C.~Bambi, K.~Freese, S.~Vagnozzi and L.~Visinelli,
\href{https://journals.aps.org/prd/abstract/10.1103/PhysRevD.100.044057}{Phys. Rev. D \textbf{100}, no.4, 044057 (2019).}

\bibitem{Jusufi:2021lei}
K.~Jusufi, S.~Kumar, M.~Azreg-A{\"\i}nou, M.~Jamil, Q.~Wu and C.~Bambi,
\href{https://link.springer.com/article/10.1140/epjc/s10052-022-10603-7}{Eur. Phys. J. C \textbf{82}, no.7, 633 (2022).}

\bibitem{Solanki:2022glc}
J.~Solanki and V.~Perlick,
\href{https://journals.aps.org/prd/abstract/10.1103/PhysRevD.105.064056}{Phys. Rev. D \textbf{105}, no.6, 064056 (2022).}

\bibitem{Mishra:2019trb}
A.~K.~Mishra, S.~Chakraborty and S.~Sarkar,
\href{https://journals.aps.org/prd/abstract/10.1103/PhysRevD.99.104080}{Phys. Rev. D \textbf{99}, no.10, 104080 (2019).}

\bibitem{Ghosh:2023kge}
R.~Ghosh, S.~Sk and S.~Sarkar,
\href{https://journals.aps.org/prd/abstract/10.1103/PhysRevD.108.L041501}{Phys. Rev. D \textbf{108}, no.4, 4 (2023).}

\bibitem{Banerjee:2019nnj}
I.~Banerjee, S.~Chakraborty and S.~SenGupta,
\href{https://journals.aps.org/prd/abstract/10.1103/PhysRevD.101.041301}{Phys. Rev. D \textbf{101}, no.4, 041301 (2020).}

\bibitem{Banerjee:2022jog}
I.~Banerjee, S.~Chakraborty and S.~SenGupta,
\href{https://journals.aps.org/prd/abstract/10.1103/PhysRevD.106.084051}{Phys. Rev. D \textbf{106}, no.8, 084051 (2022).}

\bibitem{Cardoso:2014sna}
V.~Cardoso, L.~C.~B.~Crispino, C.~F.~B.~Macedo, H.~Okawa and P.~Pani,
\href{https://journals.aps.org/prd/abstract/10.1103/PhysRevD.90.044069}{Phys. Rev. D \textbf{90}, no.4, 044069 (2014).}

\bibitem{Chen:2022kzv}
Y.~Chen, X.~Xue, R.~Brito and V.~Cardoso,
\href{https://journals.aps.org/prl/abstract/10.1103/PhysRevLett.130.111401}{Phys. Rev. Lett. \textbf{130}, no.11, 111401 (2023).}

\bibitem{Ozel:2021ayr}
F.~Ozel, D.~Psaltis and Z.~Younsi,
\href{https://iopscience.iop.org/article/10.3847/1538-4357/ac9fcb}{Astrophys. J. \textbf{941}, no.1, 88 (2022).}

\bibitem{Qiao:2025ojr}
C.~K.~Qiao, P.~Su and Y.~Huang,
\href{https://link.springer.com/article/10.1140/epjc/s10052-025-14435-z}{Eur. Phys. J. C \textbf{85}, no.6, 709 (2025).}

\bibitem{Faggert:2025eja}
J.~C.~Faggert, F.~Ozel and D.~Psaltis,
[arXiv:2506.15783 [astro-ph.HE]].

\bibitem{Suzuki:2025ptr}
H.~Suzuki, Y.~Mizuno, A.~Uniyal, I.~K.~Dihingia, T.~Nguyen and C.~k.~Chan,
[arXiv:2511.20756 [astro-ph.HE]].

\bibitem{Uniyal:2025uvc}
A.~Uniyal, I.~K.~Dihingia, Y.~Mizuno and L.~Rezzolla,
\href{https://www.nature.com/articles/s41550-025-02695-4}{Nature Astron. \textbf{1}, 8 (2025).}

\bibitem{Uniyal:2022vdu}
A.~Uniyal, R.~C.~Pantig and A.~{\"O}vg{\"u}n,
\href{https://www.sciencedirect.com/science/article/abs/pii/S2212686423000122?via%3Dihub}{Phys. Dark Univ. \textbf{40}, 101178 (2023).}

\bibitem{Saurabh:2020zqg}
K.~Saurabh and K.~Jusufi,
\href{https://link.springer.com/article/10.1140/epjc/s10052-021-09280-9}{Eur. Phys. J. C \textbf{81}, no.6, 490 (2021).}

\bibitem{EventHorizonTelescope:2025uqi}
Saurabh \textit{et al.} [Event Horizon Telescope],
\href{https://arxiv.org/abs/2512.08970}{[arXiv:2512.08970 [astro-ph.HE]].}

\bibitem{Tiede:2022grp}
P.~Tiede, M.~D.~Johnson, D.~W.~Pesce, D.~C.~M.~Palumbo, D.~O.~Chang and P.~Galison,
\href{https://www.mdpi.com/2075-4434/10/6/111}{Galaxies \textbf{10}, no.6, 111.} (2022)

\bibitem{Broderick:2022tfu}
A.~E.~Broderick, D.~W.~Pesce, P.~Tiede, H.~Y.~Pu, R.~Gold, R.~Anantua, S.~Britzen, C.~Ceccobello, K.~Chatterjee and Y.~Chen, \textit{et al.}
\href{https://iopscience.iop.org/article/10.3847/1538-4357/ac7c1d}{Astrophys. J. \textbf{935}, 61 (2022).}

\bibitem{Lockhart:2022rui}
W.~Lockhart and S.~E.~Gralla,
\href{https://academic.oup.com/mnras/article/517/2/2462/6726640?login=true}{Mon. Not. Roy. Astron. Soc. \textbf{517}, no.2, 2462-2470 (2022).}

\bibitem{lensing1}
S.~Paul,
\href{doi:10.1103/PhysRevD.102.064045}{Phys. Rev. D \textbf{102}, no.6, 064045 (2020).}

\bibitem{Virbhadra:2022iiy}
K.~S.~Virbhadra,
\href{https://journals.aps.org/prd/abstract/10.1103/PhysRevD.106.064038}{Phys. Rev. D \textbf{106}, no.6, 064038 (2022).}

\bibitem{Virbhadra:2008ws}
K.~S.~Virbhadra,
\href{https://journals.aps.org/prd/abstract/10.1103/PhysRevD.79.083004}{Phys. Rev. D \textbf{79}, 083004 (2009).}

\bibitem{Virbhadra:2002ju}
K.~S.~Virbhadra and G.~F.~R.~Ellis,
\href{https://journals.aps.org/prd/abstract/10.1103/PhysRevD.65.103004}{Phys. Rev. D \textbf{65}, 103004 (2002).}

\bibitem{Virbhadra:1999nm}
K.~S.~Virbhadra and G.~F.~R.~Ellis,
\href{https://journals.aps.org/prd/abstract/10.1103/PhysRevD.62.084003}{Phys. Rev. D \textbf{62}, 084003 (2000).}

\bibitem{Bozza:2010xqn}
V.~Bozza,
\href{https://link.springer.com/article/10.1007/s10714-010-0988-2}{Gen. Rel. Grav. \textbf{42}, 2269-2300 (2010).}

\bibitem{Chen:2023uuy}
D.~Chen, Y.~Chen, P.~Wang, T.~Wu and H.~Wu,
\href{https://link.springer.com/article/10.1140/epjc/s10052-024-12950-z}{Eur. Phys. J. C \textbf{84}, no.6, 584 (2024).}

\bibitem{Khodadi:2020gns}
M.~Khodadi and E.~N.~Saridakis,
\href{https://www.sciencedirect.com/science/article/abs/pii/S2212686421000662?via%3Dihub}{Phys. Dark Univ. \textbf{32}, 100835 (2021).}

\bibitem{Saurabh:2022jjv}
Saurabh, P.~Bambhaniya and P.~S.~Joshi,
\href{[arXiv:2202.00588 [gr-qc]]}{https://arxiv.org/abs/2202.00588}

\bibitem{Chen:2022nbb}
Y.~Chen, R.~Roy, S.~Vagnozzi and L.~Visinelli,
\href{https://journals.aps.org/prd/abstract/10.1103/PhysRevD.106.043021}{Phys. Rev. D \textbf{106}, no.4, 043021 (2022).}

\bibitem{Vagnozzi:2022moj}
S.~Vagnozzi, R.~Roy, Y.~D.~Tsai, L.~Visinelli, M.~Afrin, A.~Allahyari, P.~Bambhaniya, D.~Dey, S.~G.~Ghosh and P.~S.~Joshi, \textit{et al.}
\href{https://iopscience.iop.org/article/10.1088/1361-6382/acd97b}{Class. Quant. Grav. \textbf{40}, no.16, 165007 (2023).}

\bibitem{Khodadi:2022pqh}
M.~Khodadi and G.~Lambiase,
\href{https://journals.aps.org/prd/abstract/10.1103/PhysRevD.106.104050}{Phys. Rev. D \textbf{106}, no.10, 104050 (2022).}

\bibitem{Khodadi:2021gbc}
M.~Khodadi, G.~Lambiase and D.~F.~Mota,
\href{https://iopscience.iop.org/article/10.1088/1475-7516/2021/09/028}{JCAP \textbf{09}, 028 (2021).}

\bibitem{KumarWalia:2022ddq}
R.~Kumar Walia,
\href{https://iopscience.iop.org/article/10.1088/1475-7516/2023/03/029}{JCAP \textbf{03}, 029 (2023).}

\bibitem{Eisenhauer:2005cv} 
  F. Eisenhauer {\it et al.},
\href{https://iopscience.iop.org/article/10.1086/430667/meta}{  Astrophys. J.  {\bf 628}, 246 (2005).}

\bibitem{center1}
S. Gillessen {\it et al.},
\href{https://iopscience.iop.org/article/10.3847/1538-4357/aa5c41/meta}{Astrophys. J.  {\bf 837}, 30 (2017). }

\bibitem{Wu:2023wld}
M.~H.~Wu, H.~Guo and X.~M.~Kuang,
\href{https://journals.aps.org/prd/abstract/10.1103/PhysRevD.107.064033}{Phys. Rev. D \textbf{107}, no.6, 064033 (2023).}

\bibitem{Deng:2020yfm}
X.~M.~Deng,
\href{doi:10.1016/j.dark.2020.100629}{Phys. Dark Univ. \textbf{30}, 100629 (2020).}

\bibitem{lensing2} 
  S. Sahu, M. Patil, D. Narasimha, and P. S. Joshi,
\href{https://journals.aps.org/prd/abstract/10.1103/PhysRevD.86.063010}{Phys. Rev. D 86, 063010 – Published 19 September 2012.}

\bibitem{lensing3} 
 S. Sahu, M. Patil, D. Narasimha, and P. S. Joshi,
\href{https://journals.aps.org/prd/abstract/10.1103/PhysRevD.88.103002}{Phys. Rev. D 88, 103002 – Published 6 November 2013.}

\bibitem{Patel:2022vlu}
V.~Patel, D.~Tahelyani, A.~B.~Joshi, D.~Dey and P.~S.~Joshi,
\href{https://link.springer.com/article/10.1140/epjc/s10052-022-10638-w}{Eur. Phys. J. C \textbf{82} (2022) no.9, 798.}

\bibitem{lensing6}
N.~Tsukamoto,
\href{https://journals.aps.org/prd/abstract/10.1103/PhysRevD.102.104029}{Phys. Rev. D \textbf{102}, no.10, 104029 (2020).}

\bibitem{Gao:2020wjz}
B.~Gao and X.~M.~Deng,
\href{doi:10.1016/j.aop.2020.168194}{Annals Phys. \textbf{418}, 168194 (2020).}

\bibitem{Sajadi:2023ybm}
S.~N.~Sajadi, M.~Khodadi, O.~Luongo and H.~Quevedo,
\href{https://www.sciencedirect.com/science/article/abs/pii/S2212686424001079}{Phys. Dark Univ. \textbf{45}, 101525 (2024).}

\bibitem{Joshi:2020tlq}
A.~B.~Joshi, D.~Dey, P.~S.~Joshi and P.~Bambhaniya,
\href{doi:10.1103/PhysRevD.102.024022}{Phys. Rev. D \textbf{102}, no.2, 024022 (2020).}

\bibitem{Joshi:2023ugm}
A.~B.~Joshi, K.~Mosani and P.~S.~Joshi,
\href{https://journals.aps.org/prd/abstract/10.1103/PhysRevD.109.064019}{Phys. Rev. D \textbf{109}, no.6, 064019 (2024).}

\bibitem{Dey:2020bgo}
D.~Dey, R.~Shaikh and P.~S.~Joshi,
\href{doi:10.1103/PhysRevD.103.024015}{Phys. Rev. D \textbf{103}, no.2, 024015 (2021).}

\bibitem{Kaur}
K.~P.~Kaur, P.~S.~Joshi, D.~Dey, A.~B.~Joshi and R.~P.~Desai,
\href{https://arxiv.org/abs/2106.13175}{arXiv:2106.13175 [gr-qc].}

\bibitem{Khodadi:2024ubi}
M.~Khodadi, S.~Vagnozzi and J.~T.~Firouzjaee,
\href{https://www.nature.com/articles/s41598-024-78264-y}{Sci. Rep. \textbf{14}, no.1, 26932 (2024).}

\bibitem{Viththani:2024fod}
D.~P.~Viththani, A.~B.~Joshi, T.~Bhanja and P.~S.~Joshi,
\href{https://link.springer.com/article/10.1140/epjc/s10052-024-12746-1}{Eur. Phys. J. C \textbf{84}, no.4, 383 (2024).}

\bibitem{Uniyal:2025hik}
A.~Uniyal, I.~K.~Dihingia, Y.~Mizuno and W.~Klu{\'z}niak,
\href{https://iopscience.iop.org/article/10.3847/1538-4357/ae071a}{Astrophys. J. \textbf{993}, no.1, 97 (2025).}

\bibitem{Tahelyani:2022uxw}
D.~Tahelyani, A.~B.~Joshi, D.~Dey and P.~S.~Joshi,
\href{https://arxiv.org/pdf/2205.04055.pdf}{arXiv:2205.04055 [gr-qc], (2022).}

\bibitem{Olivares-Sanchez:2024dfh}
H.~R.~Olivares-S{\'a}nchez, P.~Kocherlakota and C.~A.~R.~Herdeiro,
``GRMHD Simulations of~Accretion Onto Exotic Compact Objects,'' in New Frontiers in GRMHD Simulations, edited by Cosimo Bambi, Yosuke Mizuno, Swarnim Shashank, and Feng Yuan \href{https://link.springer.com/chapter/10.1007/978-981-97-8522-3_15#citeas}{Springer, Singapore, 2025.}

\bibitem{Pugliese:2024bhh}
D.~Pugliese and Z.~Stuchlik,
\href{https://epjc.epj.org/articles/epjc/abs/2024/02/10052_2024_Article_12512/10052_2024_Article_12512.html}{Eur. Phys. J. C \textbf{84}, no.2, 158 (2024).}

\bibitem{Patra:2023epx}
S.~Patra, B.~R.~Majhi and S.~Das,
\href{https://iopscience.iop.org/article/10.1088/1475-7516/2024/01/060}{JCAP \textbf{01}, 060 (2024).}

\bibitem{Prada-Mendez:2023psi}
G.~D.~Prada-M{\'e}ndez, F.~D.~Lora-Clavijo and J.~M.~Vel{\'a}squez-Cadavid,
\href{https://iopscience.iop.org/article/10.1088/1361-6382/acf17e/meta}{Class. Quant. Grav. \textbf{40}, no.19, 195011 (2023).}

\bibitem{Pugliese:2023kwp}
D.~Pugliese and Z.~Stuchl{\'\i}k,
\href{https://link.springer.com/article/10.1140/epjc/s10052-023-11305-4}{Eur. Phys. J. C \textbf{83}, no.3, 242 (2023).} erratum: \href{https://link.springer.com/article/10.1140/epjc/s10052-023-11426-w}{Eur. Phys. J. C \textbf{83}, no.4, 303 (2023).}

\bibitem{Boshkayev:2022vlv}
K.~Boshkayev, T.~Konysbayev, Y.~Kurmanov, O.~Luongo and D.~Malafarina,
\href{https://iopscience.iop.org/article/10.3847/1538-4357/ac8804}{Astrophys. J. \textbf{936}, no.2, 96 (2022).}

\bibitem{Gan:2021xdl}
Q.~Gan, P.~Wang, H.~Wu and H.~Yang,
\href{https://journals.aps.org/prd/abstract/10.1103/PhysRevD.104.044049}{Phys. Rev. D \textbf{104}, no.4, 044049 (2021).}

\bibitem{Hu:2020usx}
Z.~Hu, Z.~Zhong, P.~C.~Li, M.~Guo and B.~Chen,
\href{https://journals.aps.org/prd/abstract/10.1103/PhysRevD.103.044057}{Phys. Rev. D \textbf{103}, no.4, 044057 (2021).}

\bibitem{Huang:2024bbs}
H.~Huang, J.~Kunz and D.~Mitra,
\href{https://iopscience.iop.org/article/10.1088/1475-7516/2024/05/007}{JCAP \textbf{05}, 007 (2024).}

\bibitem{Feng:2022evy}
J.~C.~Feng, S.~Chakraborty and V.~Cardoso,
\href{https://journals.aps.org/prd/abstract/10.1103/PhysRevD.107.044050}{Phys. Rev. D \textbf{107}, no.4, 044050 (2023).}

\bibitem{DiFilippo:2024poc}
F.~Di Filippo and L.~Rezzolla,
\href{https://journals.aps.org/prd/abstract/10.1103/PhysRevD.111.L021504}{Phys. Rev. D \textbf{111}, no.2, 2 (2025).}

\bibitem{Bambi:2008jg}
C.~Bambi and K.~Freese,
\href{https://journals.aps.org/prd/abstract/10.1103/PhysRevD.79.043002}{Phys. Rev. D \textbf{79}, 043002 (2009).}

\bibitem{Bambi:2017iyh}
C.~Bambi,
\href{https://onlinelibrary.wiley.com/doi/10.1002/andp.201700430}{Annalen Phys. \textbf{530}, 1700430 (2018).}

\bibitem{grava2} 
  Ken-ichi Nakao, Chul-Moon Yoo, and Tomohiro Harada
\href{https://journals.aps.org/prd/abstract/10.1103/PhysRevD.99.044027}{Phys.\ Rev.\ D {\bf 99}, 044027 (2019).} 

\bibitem{grava3} 
  Nobuyuki Sakai, Hiromi Saida, and Takashi Tamaki
\href{https://journals.aps.org/prd/abstract/10.1103/PhysRevD.90.104013}{Phys.\ Rev.\ D {\bf 90}, 104013 (2014).} 

\bibitem{bambi} 
  Cosimo Bambi
\href{https://journals.aps.org/prd/abstract/10.1103/PhysRevD.87.107501}{Phys. Rev. D 87, 107501 – Published 3 May 2013.} 

\bibitem{ohgami:2015}
T. Ohgami and N. Sakai, 
\href{https://journals.aps.org/prd/abstract/10.1103/PhysRevD.91.124020}{Phys. Rev. D {\bf 91}, 124020 (2015).}

\bibitem{Huang:2023yqd}
H.~Huang, J.~Kunz, J.~Yang and C.~Zhang,
\href{https://journals.aps.org/prd/abstract/10.1103/PhysRevD.107.104060}{Phys. Rev. D \textbf{107}, no.10, 104060 (2023).}

\bibitem{Shaikh:2022ivr}
R.~Shaikh,
\href{https://academic.oup.com/mnras/article/523/1/375/7157137}{Mon. Not. Roy. Astron. Soc. \textbf{523}, no.1, 375-384 (2023).}

\bibitem{Zhu:2021tgb}
Y.~Zhu and T.~Wang,
\href{https://journals.aps.org/prd/abstract/10.1103/PhysRevD.104.104052}{Phys. Rev. D \textbf{104}, no.10, 104052 (2021).}

\bibitem{Mustapha1}
Azreg-Aïnou, M.,
\href{https://doi.org/10.1140/epjc/s10052-014-2865-8}{Eur. Phys. J. C {\bf 74}, 2865 (2014).}

\bibitem{Cunha:2020azh}
P.~V.~P.~Cunha and C.~A.~R.~Herdeiro,
\href{https://journals.aps.org/prl/abstract/10.1103/PhysRevLett.124.181101}{Phys. Rev. Lett. \textbf{124}, no.18, 181101 (2020).}

\bibitem{Ghosh:2021txu}
R.~Ghosh and S.~Sarkar,
\href{https://journals.aps.org/prd/abstract/10.1103/PhysRevD.104.044019}{Phys. Rev. D \textbf{104}, no.4, 044019 (2021).}

\bibitem{Solanki:2021mkt}
D.~N.~Solanki, P.~Bambhaniya, D.~Dey, P.~S.~Joshi and K.~N.~Pathak,
\href{https://link.springer.com/article/10.1140/epjc/s10052-022-10045-1}{Eur. Phys. J. C \textbf{82}, no.1, 77 (2022).}

\bibitem{Shaikh:2019hbm}
R.~Shaikh and P.~S.~Joshi,
\href{https://iopscience.iop.org/article/10.1088/1475-7516/2019/10/064}{JCAP \textbf{10}, 064 (2019).}

\bibitem{Bardeen}
J. ~M. Bardeen, \textit{Non-singular general relativistic gravitational collapse}, in Proceedings of the International Conference GR5, Tbilisi, U.S.S.R. (1968).

\bibitem{Bardeen:2014uaa}
J.~M.~Bardeen,
\href{https://arxiv.org/abs/1406.4098}{[arXiv:1406.4098 [gr-qc]].}

\bibitem{Roman:1983zza}
T.~A.~Roman and P.~G.~Bergmann,
\href{https://journals.aps.org/prd/abstract/10.1103/PhysRevD.28.1265}{Phys. Rev. D \textbf{28}, 1265-1277 (1983).}

\bibitem{Frolov:2016pav}
V.~P.~Frolov,
\href{https://journals.aps.org/prd/abstract/10.1103/PhysRevD.94.104056}{Phys. Rev. D \textbf{94}, no.10, 104056 (2016).}

\bibitem{Hayward:2005gi}
S.~A.~Hayward,
\href{https://journals.aps.org/prl/abstract/10.1103/PhysRevLett.96.031103}{Phys. Rev. Lett. \textbf{96}, 031103 (2006).}

\bibitem{stuchlik_2019} 
Z. Stuchlik and J. Schee, 
\href{https://link.springer.com/article/10.1140/epjc/s10052-019-6543-8}{Eur. Phys. J. C {\bf 79}, 44 (2019).}

\bibitem{Bambhaniya:2021ugr}
P.~Bambhaniya, S.~K, K.~Jusufi and P.~S.~Joshi,
\href{https://journals.aps.org/prd/abstract/10.1103/PhysRevD.105.023021}{Phys. Rev. D \textbf{105}, no.2, 023021 (2022).}

\bibitem{Olmo:2023lil}
G.~J.~Olmo, J.~L.~Rosa, D.~Rubiera-Garcia and D.~Saez-Chillon Gomez,
\href{https://iopscience.iop.org/article/10.1088/1361-6382/aceacd}{Class. Quant. Grav. \textbf{40}, no.17, 174002 (2023).}

\bibitem{Boshkayev:2023fft}
K.~Boshkayev, T.~Konysbayev, Y.~Kurmanov, O.~Luongo, M.~Muccino, A.~Taukenova and A.~Urazalina,
\href{https://link.springer.com/article/10.1140/epjc/s10052-024-12446-w}{Eur. Phys. J. C \textbf{84}, no.3, 230 (2024).}

\bibitem{Kumar:2020ltt}
R.~Kumar and S.~G.~Ghosh,
\href{https://iopscience.iop.org/article/10.1088/1361-6382/abdd48}{Class. Quant. Grav. \textbf{38}, no.8, 8 (2021).}

\bibitem{Bambi:2013ufa}
C.~Bambi and L.~Modesto,
\href{https://www.sciencedirect.com/science/article/abs/pii/S0370269313002505?via%3Dihub}{Phys. Lett. B \textbf{721}, 329-334 (2013).}

\bibitem{Mustapha2}
Mustapha Azreg-Aïnou,
\href{https://doi.org/10.1016/j.physletb.2014.01.041}{Phys. Lett. B {\bf 730}, 95-98 (2014).}

\bibitem{Pal:2023wqg}
K.~Pal, K.~Pal, R.~Shaikh and T.~Sarkar,
\href{https://iopscience.iop.org/article/10.1088/1475-7516/2023/11/060}{JCAP \textbf{11}, 060 (2023).}

\bibitem{Pal:2022cxb}
K.~Pal, K.~Pal, P.~Roy and T.~Sarkar,
\href{https://link.springer.com/article/10.1140/epjc/s10052-023-11558-z}{Eur. Phys. J. C \textbf{83}, no.5, 397 (2023).}

\bibitem{Pal:2024kng}
K.~Pal, K.~Pal and T.~Sarkar,
\href{https://iopscience.iop.org/article/10.1088/1475-7516/2025/01/069}{JCAP \textbf{01}, 069 (2025).}

\bibitem{Frolov:2021vbg}
V.~P.~Frolov,
\href{https://www.sif.it/riviste/sif/ncc/econtents/2022/045/02/article/13}{Nuovo Cim. C \textbf{45}, no.2, 38 (2022).}

\bibitem{Frolov:1988vj}
V.~P.~Frolov, M.~A.~Markov and V.~F.~Mukhanov,
\href{https://journals.aps.org/prd/abstract/10.1103/PhysRevD.41.383}{Phys. Rev. D \textbf{41}, 383 (1990).}

\bibitem{Modesto:2004xx}
L.~Modesto,
\href{https://journals.aps.org/prd/abstract/10.1103/PhysRevD.70.124009}{Phys. Rev. D \textbf{70}, 124009 (2004).}

\bibitem{Ashtekar:2005cj}
A.~Ashtekar and M.~Bojowald,
\href{https://iopscience.iop.org/article/10.1088/0264-9381/22/16/014}{Class. Quant. Grav. \textbf{22}, 3349-3362 (2005).}

\bibitem{Dymnikova:1992ux}
I.~Dymnikova,
\href{https://link.springer.com/article/10.1007/BF00760226}{Gen. Rel. Grav. \textbf{24}, 235-242 (1992).}

\bibitem{Bonanno:2000ep}
A.~Bonanno and M.~Reuter,
\href{https://journals.aps.org/prd/abstract/10.1103/PhysRevD.62.043008}{Phys. Rev. D \textbf{62}, 043008 (2000).}

\bibitem{Joshi:2025ozt}
V.~Joshi and A.~B.~Joshi,
\href{https://arxiv.org/abs/2512.07786}{[arXiv:2512.07786 [gr-qc]].}

\bibitem{Morris:1988cz}
M.~S.~Morris and K.~S.~Thorne,
\href{https://pubs.aip.org/aapt/ajp/article-abstract/56/5/395/1044276/Wormholes-in-spacetime-and-their-use-for?redirectedFrom=fulltext}{Am. J. Phys. \textbf{56}, 395-412 (1988).}

\bibitem{Simpson:2018tsi}
A.~Simpson and M.~Visser,
\href{https://iopscience.iop.org/article/10.1088/1475-7516/2019/02/042}{JCAP \textbf{02}, 042 (2019).}

\bibitem{Franzin:2021vnj}
E.~Franzin, S.~Liberati, J.~Mazza, A.~Simpson and M.~Visser,
\href{https://iopscience.iop.org/article/10.1088/1475-7516/2021/07/036}{JCAP \textbf{07}, 036 (2021).}

\bibitem{Guo:2021wid}
Y.~Guo and Y.~G.~Miao,
\href{https://www.sciencedirect.com/science/article/pii/S0550321322002899?via%3Dihub}{Nucl. Phys. B \textbf{983}, 115938 (2022).}

\bibitem{Kala:2025fld}
S.~Kala and J.~Singh,
\href{https://link.springer.com/article/10.1140/epjc/s10052-025-14793-8}{Eur. Phys. J. C \textbf{85}, no.9, 1047 (2025).}

\end{thebibliography}
\end{document}